\documentclass{emulateapj}


\usepackage{mathtools}
\usepackage[]{natbib}
\usepackage{graphics}
\usepackage{graphicx}

\newcommand{\per}{\ensuremath{^{-1}}}
\newcommand{\persq}{\ensuremath{^{-2}}}
\newcommand{\hal}{H\ensuremath{\alpha}}
\newcommand{\hbeta}{H\ensuremath{\beta}}

\newcommand{\hst}{\emph{HST}}
\newcommand{\msun}{\ensuremath{{M}_{\odot}}}
\newcommand{\mbh}{\ensuremath{M_\mathrm{BH}}}
\newcommand{\mNC}{\ensuremath{M_\mathrm{NC}}}
\newcommand{\mbulge}{\ensuremath{M_\mathrm{bulge}}}

\newcommand{\kms}{km~s\ensuremath{^{-1}}}

\newcommand{\sersic}{S\'{e}rsic}
\newcommand{\Reff}{\ensuremath{R_\mathrm{e}}}

\newcommand{\extinction}{\ensuremath{A} (mag)}
\newcommand{\appmag}{\ensuremath{m} (mag)}
\newcommand{\absmag}{\ensuremath{M} (mag)}
\newcommand{\reff}{\ensuremath{R_\mathrm{e}} (pc)}
\newcommand{\sersicn}{\ensuremath{n}}
\newcommand{\axisratio}{\ensuremath{b/a}}
\newcommand{\Rslope}{$d\log(\Reff/\mathrm{pc})/d\log(\lambda/\mathrm{\AA})$}
\newcommand{\qslope}{$d\log(b/a)/d\log(\lambda/\mathrm{\AA})$}
\newcommand\xput[2][0.5]{\rule{#1\linewidth}{0pt}\makebox[0pt][c]{#2}\hfill}

\begin{document}
\bibliographystyle{apj}
%\singlespacing

%\renewcommand*{\thefootnote}{\fnsymbol{footnote}}
\title{The Structure of Nuclear Star Clusters in Nearby Late-type Spiral Galaxies from Hubble Space Telescope Wide Field Camera 3 Imaging \footnotemark[*]}

\footnotetext[*]{Based on observations made with the NASA/ESA Hubble Space Telescope, obtained at the Space Telescope Science Institute, which is operated by the Association of Universities for Research in Astronomy, Inc., under NASA contract NAS 5-26555. These observations are associated with program GO-12163}

\author{
  Daniel J. Carson\altaffilmark{1}, 
  Aaron J. Barth\altaffilmark{1}, 
  Anil C. Seth\altaffilmark{2}, 
  Mark den Brok\altaffilmark{2}, 
  Michele Cappellari\altaffilmark{3}, 
  Jenny E. Greene\altaffilmark{4}, 
  Luis C. Ho\altaffilmark{5,6}, 
  Nadine Neumayer\altaffilmark{7}
}

\altaffiltext{1}{Department of Physics and Astronomy, University of California Irvine, 4129 Frederick Reines Hall, Irvine, CA 92697, USA}
\altaffiltext{2}{Department of Physics and Astronomy, University of Utah, Salt Lake City, UT 84112, USA}
\altaffiltext{3}{Sub-Department of Astrophysics, Department of Physics, University of Oxford, Denys Wilkinson Building, Keble Road, Oxford OX1 3RH, UK}
\altaffiltext{4}{Department of Astrophysical Sciences, Princeton University, Peyton Hall -- Ivy Lane, Princeton, NJ 08544, USA}
\altaffiltext{5}{Kavli Institute for Astronomy and Astrophysics, Peking University, Beijing 100871, China}
\altaffiltext{6}{Department of Astronomy, School of Physics, Peking University, Beijing 100871, China}
\altaffiltext{7}{Max-Planck-Institut f\"ur Astronomie, K\"onigstuhl 17, D-69117 Heidelberg, Germany}

\begin{abstract} 
We obtained \emph{Hubble Space Telescope}/Wide Field Camera 3 imaging of a sample of ten of the nearest and brightest nuclear clusters residing in late-type spiral galaxies, in seven bands that span the near-ultraviolet to the near-infrared. Structural properties of the clusters were measured by fitting two-dimensional surface brightness profiles to the images using \texttt{GALFIT}. The clusters exhibit a wide range of structural properties, with F814W absolute magnitudes that range from $-11.2$ mag to $-15.1$ mag and F814W effective radii that range from $1.4$ to $8.3$ pc. For six of the ten clusters in our sample, we find changes in the effective radius with wavelength, suggesting radially varying stellar populations. In four of the objects, the effective radius increases with wavelength, indicating the presence of a younger population which is more concentrated than the bulk of the stars in the cluster. However, we find a general decrease in effective radius with wavelength in two of the objects in our sample, which may indicate extended, circumnuclear star formation. We also find a general trend of increasing roundness of the clusters at longer wavelengths, as well as a correlation between the axis ratios of the NCs and their host galaxies. These observations indicate that blue disks aligned with the host galaxy plane are a common feature of nuclear clusters in late-type galaxies, but are difficult to detect in galaxies that are close to face-on. In color-color diagrams spanning the near-UV through the near-IR, most of the clusters lie far from single-burst evolutionary tracks, showing evidence for multi-age populations. Most of the clusters have integrated colors consistent with a mix of an old population ($> 1$ Gyr) and a young population ($\sim100$--$300$ Myr). The wide wavelength coverage of our data provides a sensitivity to populations with a mix of ages that would not be possible to achieve with imaging in optical bands only. The surface brightness profiles presented in this work will be used for future stellar population modeling and dynamical studies of the clusters.
\end{abstract}

\section{Introduction}
\indent \emph{Hubble Space Telescope} (\hst) imaging surveys have shown that most galaxies contain bright, compact stellar systems known as nuclear clusters (NCs) in their photometric and dynamical centers \citep{carollo1997,matthews1999,boker2002}. Similar in many respects to high-mass globular clusters (GCs), NCs typically have velocity dispersions of 15 -- 35 \kms, effective radii (or half-light radii) of a few parsecs, and dynamical masses of $10^{6}$ -- $10^{7}$ \msun\ \citep{boker1999,boker2002,walcher2005}. Unlike GCs, NCs in spiral galaxies have complex star formation histories, undergoing repeated bursts of star formation, with spectra often consistent with a continuous star formation history \citep{walcher2006,rossa2006,seth2006}. Young ($< 100$ Myr) populations are found in most spiral NCs, and there is limited evidence for complex populations in early-type nuclei as well \citep{monaco2009,seth2010,lyubenova2013}. In a recent \hst\ archival study, \citet{g&b2014} analyzed the photometric and structural properties of a sample of 228 NCs in spiral galaxies, and confirmed that most NCs are similar in size to GCs, but found that the largest and brightest NCs in spiral galaxies occupy the regime between ultra-compact dwarfs (UCDs) and the nuclei of early-type galaxies in the size-luminosity plane, which is consistent with the idea that some UCDs and GCs are the remnant nuclei of tidally disrupted spiral galaxies \citep{boker2008,mieske2013,seth2014}. They also found that the effective radius of NCs tends to be smaller when measured in bluer filters, suggesting that younger stellar populations tend to be more concentrated than the rest of the stars in NCs. We note that the Milky Way's NC also has a population of young stars concentrated in the central 0.5 pc \citep{paumard2006,lu2013,feldmeier2014}.

\indent Two main scenarios have been proposed to explain the formation of NCs: in-situ formation from the accretion of low angular momentum gas onto the galactic nucleus \citep{milos2004}, and mergers of globular clusters migrating towards the galactic nucleus due to dynamical friction \citep{lotz2001}. \citet{hartmann2011} compared simulations of the migratory/merger scenario with integral field unit (IFU) observations of NGC 4244, and found that in-situ formation accounts for at least $50 \%$ of the NC's mass. \citet{antonini2013,antonini2014} and \citet{as&cd2014} demonstrated that the migratory/merger scenario predicts correlations between the masses of NCs and global properties of their host galaxies that match observed correlations. The young populations observed in most NCs rule out the sinking of globular clusters as the sole process for NC formation (i.e., the sinking of younger star clusters or gas accretion must play some role). However, our understanding of the formation of NCs is incomplete and the reason for the ubiquity of NCs still remains a mystery.  

\indent NCs and BHs are known to coexist in galaxies of all Hubble types across a wide range of masses \citep{seth2008a}, but some NCs have no central BH down to highly constraining limits. The NC in NGC 4935, for example, contains a NC \citep{f&h2003} as well as a Type 1 active galactic nucleus (AGN) powered by accretion onto a central BH \citep{f&s1989}, with mass estimates ranging from $5 \times 10^{4}$ to $3.6 \times 10^{5}$ \msun\ \citep{peterson2005,edri2012}, while dynamical studies of the NC in M33, a galaxy with a similar mass to NGC 4395, have placed remarkably tight upper limits of $1500$ -- $3000$ \msun\ on the mass of any central BH \citep{gebhardt2001,merritt2001}. These studies indicate that at least some NCs can host a central BH, but the overall occupation fraction of BHs in NCs is unknown. The masses of both NCs \citep{ferrarese2006,w&h2006} and BHs \citep{f&m2000,gebhardt2000,h&r2004} are correlated with the masses of their host galaxy bulges. Galaxies with mass above $10^{10}$ \msun\ tend to host BHs, while less massive galaxies tend to host NCs, with a transition region for galaxies with mass $10^{8}$ -- $10^{10}$ \msun\ where both objects coexist \citep{g&s2009}. Interestingly, for galaxies known to contain both a NC and a central BH, the ratio $(\mbh + \mNC)/\mbulge$ (where \mbulge\ denotes the mass of the bulge or pseudobulge component of the host galaxy) shows less scatter than either $\mbh/\mbulge$ or $\mNC/\mbulge$ \citep{k&h2013}, suggesting a link between the building of NCs and BHs. These studies add to a growing body of evidence that NCs and BHs are both generic by-products of galaxy formation, and the growth mechanisms of NCs and BHs are somehow related \citep{ferrarese2006}. 

\indent Dynamical studies of NCs in late-type spirals have been proven useful as a method for studying BHs. Previous studies of IC 342 \citep{boker1999} and NGC 3621 \citep {barth2009} have constrained \mbh\ by using \hst\ imaging and integrated stellar velocity dispersion measurements of the NC in order to carry out Jeans modeling of the clusters. \citet{kormendy2010} and \citet{n&w2012} have also performed Jeans modeling of NCs in order to set upper limits the masses of central BHs. Adaptive optics (AO) assisted IFU observations can resolve the kinematics of the nearest and brightest NCs and increase the sensitivity to detections of BHs within the clusters. For example, \citet{seth2010} placed an upper limit of $\sim 10^5$ \msun\ for the central BH in NGC 404 from dynamical modeling of the stellar kinematics of the NC using a combination of AO-assisted near-IR IFU spectroscopy, optical spectroscopy, and \hst\ imaging. Using similar methods, \citet{hartmann2011} measured a mass of $1.1 \times 10^{7}$ \msun\ for the NC in NGC 4244, and found that if the NC hosts a central BH at all, its mass is less than 1\% of the NC mass. 

\indent Studies of NCs have mainly been limited by the angular resolution of existing telescopes. The next generation of large ground-based telescopes, such as the Thirty Meter Telescope (TMT), will be able to resolve the kinematic structure of stars within the sphere of influence of BHs with much lower mass than is currently possible (down to $\sim 10^4$ \msun), and will be crucial for future dynamical searches for IMBH within NCs \citep{do2014}. In a simulation of observations of NCs with the European Extremely Large Telescope (E-ELT), \citet{E-ELT} found that observations of NCs with telescopes that can resolve individual stars in the cluster will lead to great improvements in stellar population and age-dating studies, which are currently limited by the various degeneracies that come with fitting stellar population synthesis models to the integrated cluster light. The nearest and brightest NCs will be prime targets for the next generation of large ground-based telescopes. 

\indent We present a set of uniform, high-signal-to-noise ($S/N$), unsaturated images of NCs with wavelength coverage from the near-ultraviolet (UV) to the near-infrared (IR) for a sample of ten of the nearest late-type galaxies hosting NCs. Our data represent a substantial improvement in spatial resolution, wavelength coverage, and homogeneity over previous imaging studies of NCs. Although all of the galaxies in our sample have archival \hst\ images available from previous studies, the archival data are very heterogeneous, including images that are saturated in the core of the NC and images taken with several different cameras. The homogeneity of our data allows for a more detailed analysis of the structure and stellar populations of NCs than would be possible with archival data alone. Detailed structural information of the best resolved NCs across a wide range of wavelengths may provide clues about formation mechanisms for NCs, which are currently poorly understood. 

\indent Here, we describe the detailed two-dimensional (2D) structure of this sample of NCs in a uniform way across UV, optical, and IR bands. Surface brightness profiles, effective radii, magnitudes, and colors are presented. In future work, we will describe the stellar populations and model the stellar dynamics of the NCs in order to constrain the stellar mass and the mass of any central BH in each NC. Surface brightness profiles and stellar mass-to-light profiles of the NCs will be crucial inputs for dynamical models of the clusters.

\indent The remainder of this paper is organized as follows. Section 2 describes the NC sample, the \hst\ observations and data reduction. In Section 3, we describe surface brightness profile fits to the images of the clusters. General results from these fits, which include structural parameters and magnitudes of the NCs, trends of cluster size and roundness with wavelength, and a comparison of simple stellar population models to measured NC colors are given in Section 4. A brief discussion of each individual cluster is given in Section 5. Finally, Section 6 summarizes our most important results and describes how they will applied in future work.  

\section{Data}
\indent We obtained \hst/WFC3 images of ten of the nearest and brightest nuclear clusters in seven different filter bands, spanning the near-UV to near-IR in wavelength. The images were taken over a total of 20 \hst\ orbits for the GO-12163 observing program.

\subsection{Sample Selection} 
\indent The galaxies in our sample were selected according to the following criteria: Hubble type in the range Sc--Sm, distance $< 5$ Mpc (with one exception), and the presence of a nuclear star cluster with magnitude brighter than $V=19$ mag. The Hubble type selection criterion ensures that the galaxies in our sample have little or no bulge component. NGC 3621, at a distance of 7 Mpc, was included in the sample because it is a rare example of a nearby galaxy containing a Type 2 AGN in its nuclear cluster \citep{satyapal2007,barth2009}. Basic properties of the galaxy sample are presented in Table \ref{sample}. The morphological types, \emph{B}-band magnitudes and heliocentric radial velocities listed in this table are from \emph{The Third Reference Catalogue of Bright Galaxies} \citep{RC3} and were looked up in the NASA/IPAC Extragalactic Database \footnotemark[1]. For eight of the galaxies in our sample, the distances quoted were measured using Cepheid variables. We used tip of the red giant branch (TRGB) distances for NGC 2976 and NGC 4244, because Cepheid variable distances were not available.

\footnotetext[1]{The NASA/IPAC Extragalactic Database (NED) is operated by the Jet Propulsion Laboratory, California Institute of Technology, under contract with the National Aeronautics and Space Administration.}

\begin{deluxetable*}{lccccccc}
\renewcommand{\arraystretch}{1.5}
\tablecolumns{9}
\tablewidth{0pc}
\centering
\tablecolumns{5}
\tablewidth{0pc}
\tablecaption{The Galaxy Sample}
\tablehead{Object & \ensuremath{D} (Mpc) & $m - M$ & Reference & Morphology & \ensuremath{v_{\mathrm{r}}} (\kms) & Host Galaxy \ensuremath{M_{\mathrm{B}}} & \emph{E}(\emph{B}$-$\emph{V})}
\startdata
IC 342   & $3.3 \pm 0.3$ & $27.58 \pm 0.18$ & 1  & SAB(rs)cd & 31          & $-23.99$ & 0.480  \\
M33 (NGC 598) & $0.9 \pm 0.1$ & $24.65 \pm 0.19$ & 2  & SA(s)cd   & $-179$ & $-19.05$ & 0.036 \\
NGC 247  & $3.4 \pm 0.3$  & $27.68 \pm 0.20$ & 2 & SAB(s)d   & 156         & $-18.82$ & 0.016\\
NGC 300  & $2.0 \pm 0.3$  & $26.48 \pm 0.28$ & 2 & SA(s)d    & 144         & $-18.04$ & 0.011\\
NGC 2403 & $3.1 \pm 0.2$  & $27.43 \pm 0.15$ & 3 & SAB(s)cd  & 133         & $-19.14$ & 0.034\\
NGC 2976 & \phn$3.6 \pm 0.1$*  & $27.77 \pm 0.07$ & 4 & SAc pec   & 1       & $-17.74$ & 0.064\\
NGC 3621 & $7.3 \pm 0.2$  & $29.30 \pm 0.06$ & 3  & SA(s)d    & 730         & $-20.38$ & 0.068\\
NGC 4244 & \phn$4.3 \pm 0.1$*  & $28.16 \pm 0.07$ & 4  & SA(s)cd   & 244     & $-18.96$ & 0.018\\
NGC 4395 & $4.3 \pm 0.4$  & $28.17 \pm 0.18$ & 5  & SA(s)m    & 319         & $-17.66$ & 0.015\\
NGC 7793 & $3.4 \pm 0.1$  & $27.68 \pm 0.05$ & 6  & SA(s)d    & 227         & $-18.38$ & 0.017
\enddata
\tablerefs{Redshift-independent distances and distance moduli ($m - M$) were obtained from the following: (1) \citet{saha2002}; (2) \citet{bono2010}; (3) \citet{saha2006}; (4) \citet{jacobs2009}; (5) \citet{thim2004}; (6) \citet{piet2010}. All distances were measured using Cepheid variables except those marked with an asterisk (*), which are TRGB distances. The \emph{E}(\emph{B}$-$\emph{V}) reddening due to Galactic dust for each galaxy is from \citet{s&f2011}.\\}
\label{sample}
\end{deluxetable*}

\subsection{Description of Observations} 
\indent Each cluster was observed in seven filter bands over a total of two orbits. For each galaxy, a series of four exposures was taken in each filter band. These four exposures were offset by fractional-pixel shifts relative to each other using a four-point box dither pattern. The final images were constructed by combining these exposures, which allowed for the removal of cosmic-ray hits and bad pixels.

\indent In the UVIS channel, we used the F275W, F336W, F438W, F547M, and F814W filters. We used F547M instead of a wide-band filter for the \emph{V} band because it does not include flux from the [\ion{O}{3}], \hal, or \hbeta\ emission lines. The UVIS channel CCD has a pixel scale of 0\farcs04 pixel$^{-1}$. At a distance of 5 Mpc, a pixel on the UVIS CCD subtends 0.96 pc. The point spread functions (PSFs) range in full width at half maximum from 0\farcs092 in F275W to 0\farcs196 in F153M. These angular sizes correspond to physical sizes of 2.20 pc and 4.70 pc at a distance of 5 Mpc. Since NCs typically have effective radii of a few pc, WFC3 provides sufficient spatial resolution for studies of their structure. In order to observe each NC with several filters in just two orbits, it is necessary to reduce readout time and eliminate buffer dump overheads. To accomplish this, we used the MS512C subarray readout mode, which gives a $20'' \times 20''$ field of view. This field of view is more than wide enough to cover the entire NC and sample the region where the NC surface brightness profile merges with that of the host galaxy. 

\indent In the IR channel, we used two medium-band filters, F127M and F153M, to sample the \emph{J} and \emph{H} bands because the NCs are bright enough to saturate with very short exposure times ($<10$ seconds) using wide-band filters. The IR channel HgCdTe array has coarser resolution than the UVIS channel CCD, with a scale of 0\farcs13 pixel$^{-1}$. The IRSUB256 subarray was used, giving a $35'' \times 31''$ field of view for the IR images. We used SPARS10 sampling sequences with NSAMP=10, with the exception of the two brightest NCs, IC 342 and M33, where RAPID readout mode was used to avoid saturation. 

\begin{figure}
\xput[0.5]{\includegraphics[angle=270,scale=0.37]{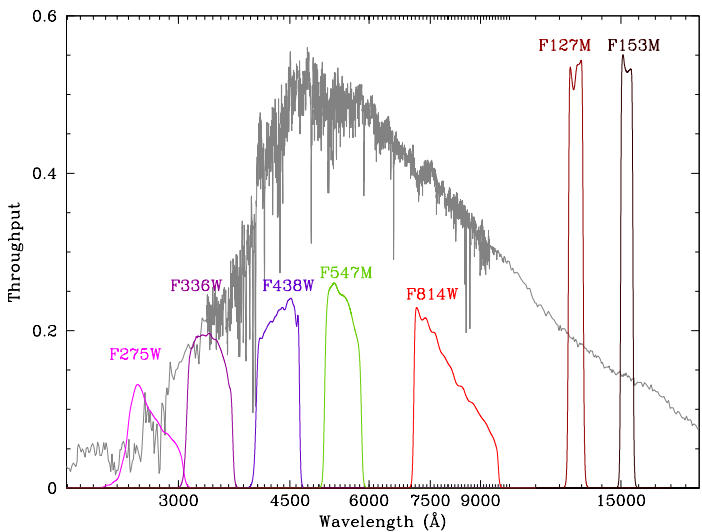}}
\caption{Throughput curves for the seven \hst/WFC3 filters taken from the PyRAF/STSDAS task \texttt{Calcband}. Also shown in gray is a model spectrum of a 1.4 Gyr old stellar population with solar metallicity from \citet{bc2003}.}
\label{throughput}
\end{figure}

\indent Throughput curves of the \hst/WFC3 filters are shown in Figure \ref{throughput}. Table \ref{filters} lists the central wavelengths and band widths of the filters as well as the exposure times used. Shorter exposure times were used for the two galaxies with the brightest NCs, IC 342 and M33, in order to avoid saturation. Due to the shorter exposure times and RAPID sampling, the sequence of 4 exposures was duplicated for the IR observations of IC 342 and M33.

\indent We obtained the reddening \emph{E}(\emph{B}$-$\emph{V}) for each galaxy from the latest Galactic dust maps \citep{s&f2011}, and we calculated the extinction due to Galactic dust in each band for all of the galaxies in the sample using the PyRAF/STSDAS task \texttt{Calcphot} \citep{synphot}. We applied the appropriate reddening to an F star template spectrum using the \citet{c&c&m1989} reddening law. From simulated \hst/WFC3 observations, we obtained the magnitudes of the reddened and unreddened template star in each band, and determined the extinction from the difference between the two. All absolute magnitudes listed in this paper have been corrected for Galactic extinction and are given in the Vega magnitude system. The conversion between magnitudes in the Vega system and magnitudes in the AB system ($m_{\mathrm{AB}} - m_{\mathrm{Vega}}$) is 1.50 mag in F275W, 1.18 mag in F336W, $-0.15$ mag in F438W, 0.42 in F814W, 0.96 in F127M, and 1.25 mag in F153M.

\begin{deluxetable*}{ccccccc}
\centering
\tablecolumns{4}
\tablewidth{0pc}
\tablecaption{\hst/WFC3 Filter Properties \& Exposure Times}
\tablehead{
& & \colhead{Central} & \colhead{Band} & \colhead{Exp. Time} & \colhead{IC 342 Exp.} & \colhead{M33 Exp.}\\
\colhead{Filter} & \colhead{Channel} & \colhead{Wavelength (\AA)} & \colhead{Width (\AA)} & \colhead{(s)} & \colhead{Time (s)} & \colhead{Time (s)}}  
\startdata
F275W    & UVIS & 2710.2\phn\phn\phn  & 164.51\phn & \phn\phn\phn$4 \times 380$  & \phn\phn\phn$4 \times 130$  & \phn\phn\phn$4 \times 130$ \\
F336W    & UVIS & 3354.8\phn\phn\phn  & 158.44\phn & \phn\phn\phn$4 \times 290$  & \phn\phn$4 \times 55$  & \phn\phn$4 \times 55$  \\
F438W    & UVIS & 4326.5\phn\phn\phn  & 197.30\phn & \phn\phn\phn$4 \times 115$  & \phn\phn$4 \times 35$  & \phn\phn$4 \times 42$ \\
F547M    & UVIS & 5447.4\phn\phn\phn  & 206.22\phn & \phn\phn$4 \times 95$  & \phn\phn$4 \times 25$ & \phn\phn$4 \times 90$ \\
F814W    & UVIS & 8029.5\phn\phn\phn  & 663.33\phn & \phn\phn$4 \times 75$ & \phn\phn$4 \times 15$ & \phn\phn$4 \times 40$ \\
F127M    & IR   & 12740.0\phn\phn\phn\phn & 249.55\phn & \phn\phn$4 \times 60$  & \phn$8 \times 5$ & \phn$8 \times 5$ \\
F153M    & IR   & 15322.0\phn\phn\phn\phn & 378.94\phn & \phn\phn$4 \times 60$  & \phn$8 \times 5$  & \phn$8 \times 5$
\enddata
\tablecomments{The central wavelength listed is the ``pivot wavelength'', a source-independent measure of the characteristic wavelength of the filter, and the band width is the ``passband rectangular width'', the integral with respect to wavelength of the throughput across the filter passband divided by the maximum throughput, as defined in the WFC3 Instrument Handbook \citep{wfc3}.} 
\label{filters}
\end{deluxetable*}

\subsection{Data Reduction}
\begin{figure*}
\centering
\includegraphics[scale=1.4]{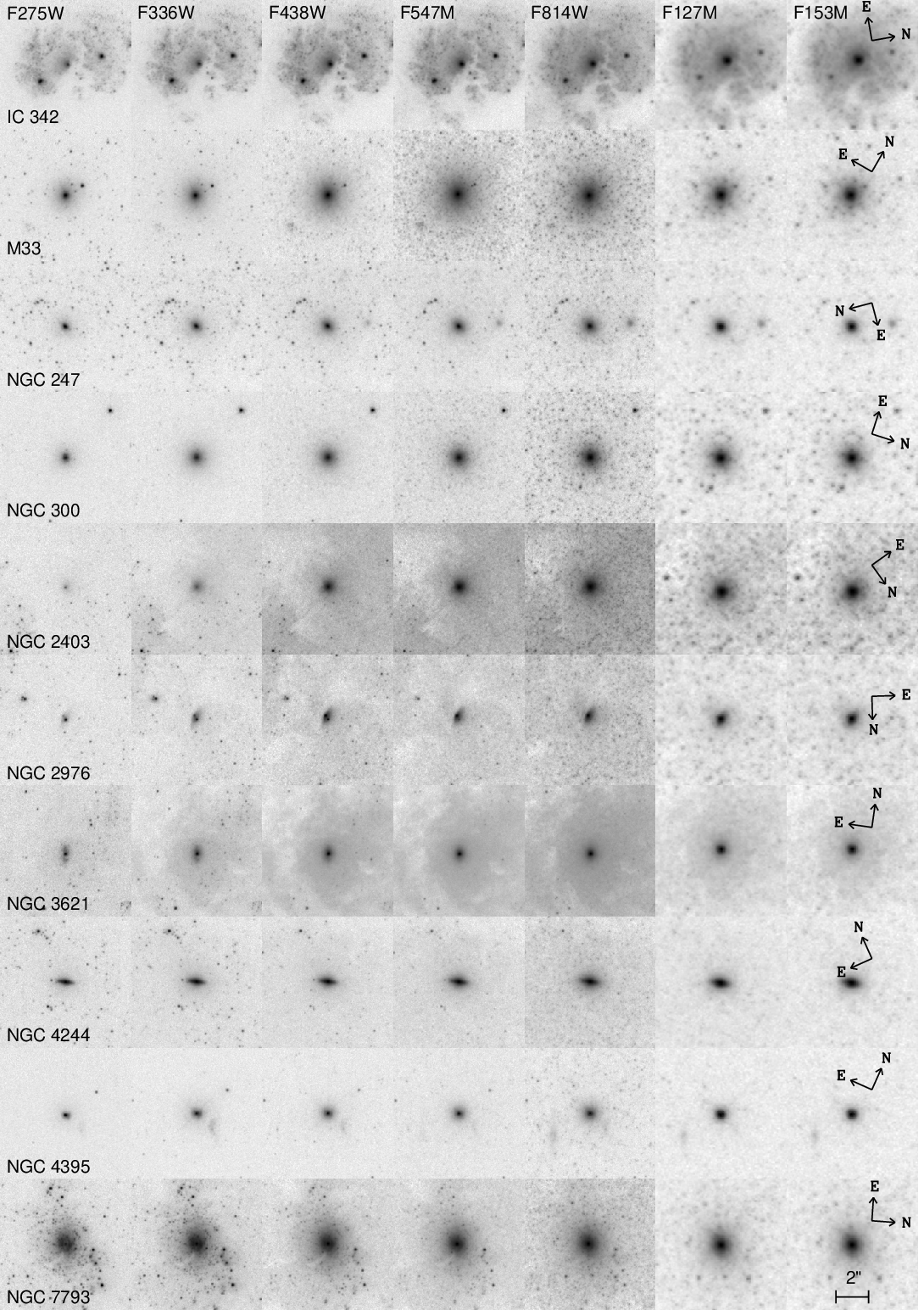}
\caption{Inverted grayscale images of the inner $8'' \times 8''$ of each galaxy in all seven \hst/WFC3 filters, shown with an asinh stretch. Images are displayed with their native orientation.}
\label{all_images}
\end{figure*}

\begin{figure*}
\centering
\includegraphics[scale=1.6]{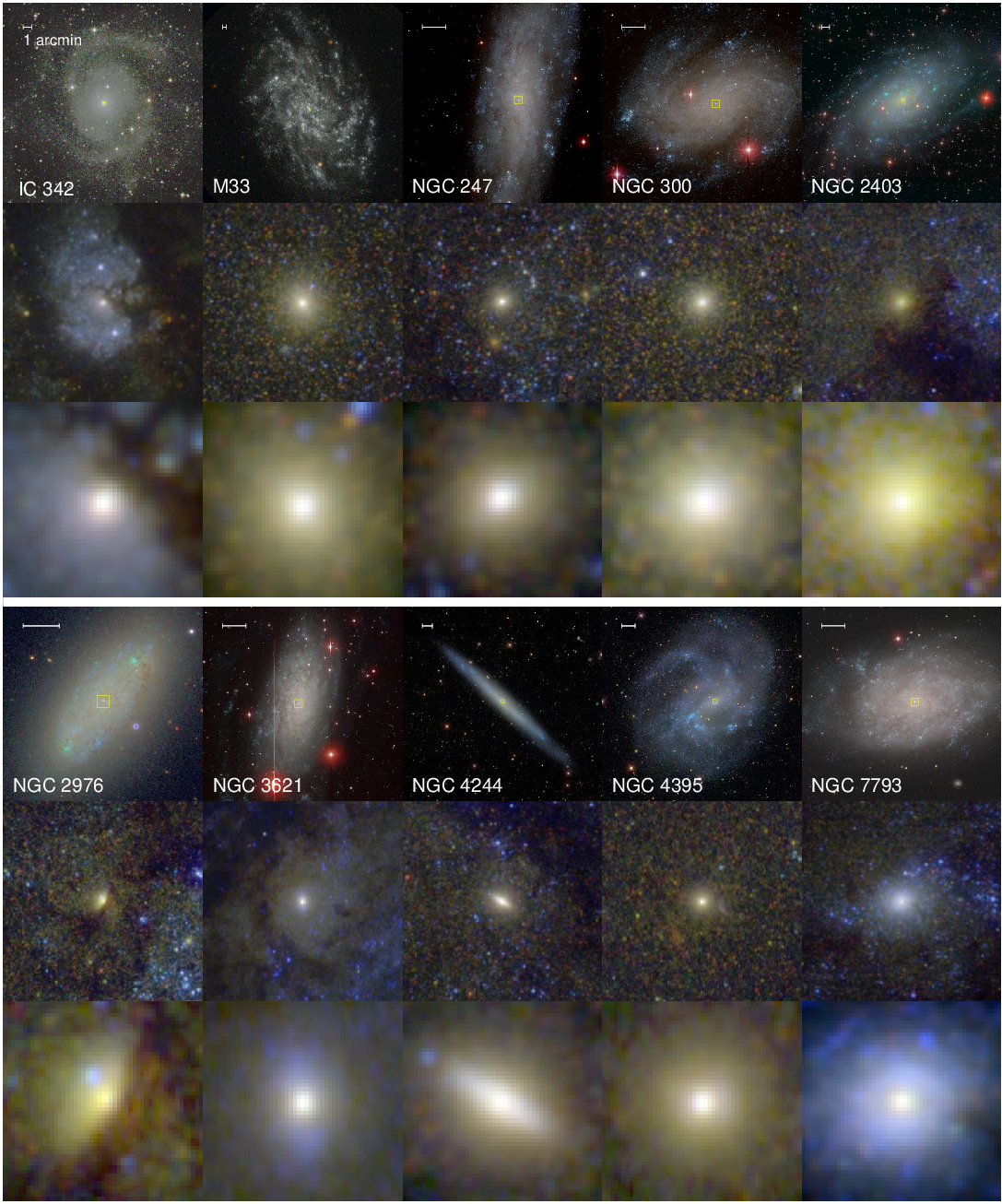}
\caption{Color composite images of each NC and its host galaxy. For each object, the top panel shows the entire host galaxy, the middle panel shows the inner $12'' \times 12''$ of the galaxy, and the bottom panel shows the inner $2'' \times 2''$ centered on the NC. Host galaxy images were taken from the following sources: 2MASS \citep[IC 342]{2MASSa}, GALEX \citep[M33]{gildepaz2007}, SDSS \citep[NGC 2403, NGC 2976, NGC 4244, and NGC 4395]{SDSS}, and CGS \citep[NGC 247, NGC 300, NGC 3621, and NGC 7793]{CGS}. For the NC images, the \hst/WFC3 F275W, F547M, and F814W images were used to populate the blue, green and red color channels, respectively. The yellow boxes on the host galaxy images show the full $20'' \times 20''$ UVIS field of view and the scale bar denotes $1'$. All of the images have been rotated so that north is up and are shown with an asinh stretch.}
\label{master_color_images}
\end{figure*}

\indent The standard \hst\ calibration pipeline was used for flat fielding, bias correction and dark current subtraction. In order to reject cosmic-ray hits and bad pixels and to correct for the geometric distortion introduced by the optics of \hst\ and WFC3, the four offset exposures for each filter were combined into a final image using the PyRAF/STSDAS task \texttt{AstroDrizzle}, using a square kernel with \emph{pixfrac} $= 1$, exposure time weighting, and sky subtraction turned off. In addition, as part of the \texttt{AstroDrizzle} process, we resampled the IR images onto the UVIS grid for a uniform pixel size of 0\farcs04 in all bands. The resulting images are shown in Figure \ref{all_images}. Color composite images of each cluster, along with images of their host galaxies, are shown in Figure \ref{master_color_images}.

\section{Surface Brightness Profile Fitting}  
\indent A 2D model of the surface brightness profile was fit to each image using \texttt{GALFIT} version 3.0.4 \citep{peng2002,peng2010}. For each image, we specified the components of the model surface brightness profile to fit to the image (e.g., \sersic, exponential) and supplied initial estimates for the model parameters (e.g., total magnitude, position of the center of the profile). A model image based on these parameters is produced and then convolved with the corresponding \hst/WFC3 PSF. The $\chi^{2}$ difference between the \hst\ image and model image is computed, and minimization of $\chi^{2}$ is achieved (and thus, best-fitting model parameters are found) using the Levenberg-Marquardt downhill-gradient method. To properly weight each pixel in the $\chi^{2}$ calculation, we used the sigma image generated internally by GALFIT based on each input image. 

\indent We fit 2D surface brightness profiles to the images rather than fitting 1D radial profiles to elliptical isophotes of the images, because it allowed us to more accurately disentangle the instrumental PSF from the intrinsic profile of the clusters. The clusters are quite compact and the fits are therefore very sensitive to the details of the PSF profile. \hst/WFC3 PSFs have non-axisymmetric features that cannot be described by a 1D radial profile. In addition, we find that some clusters have strongly non-axisymmetric features such as flattened disk components, which cannot be modeled accurately in 1D. PSF convolution in 1D is therefore not a suitable alternative to 2D convolution for the purpose of extracting the intrinsic 2D structure of the NCs from the \hst\ images.

\subsection{PSF Models}
\begin{figure*}
\centering
\includegraphics[scale=1.6]{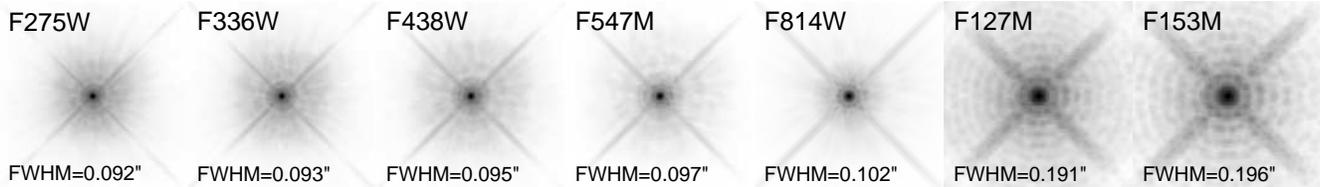}
\caption{Model PSFs in each filter. The full width at half maximum of each PSF is given in arcseconds.}
\label{PSF_images}
\end{figure*}

\indent Simulated images of the \hst/WFC3 PSFs in each filter were generated using version 7.4 of the \texttt{TinyTim} software package \citep{tinytim,tinytim2}, which creates a model \hst\ PSF based on the based on the instrument, detector chip, detector chip position, and filter used in the observations. To ensure that the model PSFs were processed in the same way as the \hst\ images, we produced four versions of each PSF on a subsampled grid with sub-pixel offsets, using the same four-point box dither pattern as the \hst/WFC3 exposures. Each model PSF was convolved with the appropriate charge diffusion kernel in order to account for the effect of electrons leaking into neighboring pixels on the CCD. The PSFs were then combined and resampled onto a final grid with a pixel size of 0\farcs04  in all bands using \texttt{AstroDrizzle}, with the settings described in Section 2.3. The resulting model PSFs are shown in Figure \ref{PSF_images}. 

\subsection{Fitting Method}
\indent The NCs sit at the center of their host galaxies, whose surface brightness profiles can generally be described by an exponential disk model because they have either a weak bulge component or no bulge component at all. The disk of the host galaxy contributes a significant amount of flux in our images, particularly in the outskirts of the NCs, and can have a strong effect on NC fit parameters such as the \sersic\ index and effective radius. Our images have a small field of view and only sample the very inner regions of the galaxies. For example, the most distant galaxy in our sample, NGC 3621, has apparent angular dimensions $12' \times 7'$ \citep{gildepaz2007}, compared to the $20'' \times 20''$ UVIS field of view. Although the NCs sit against a non-uniform background, it is not always possible to detect a gradient in the background level across the image due to the small field of view. In order to model the host galaxy surface brightness profile, we masked out the NC as well as any prominent dust lanes and attempted to fit a model consisting of an exponential disk and a flat sky level to the host galaxy. If this fit converged, we performed the NC fit with the host galaxy exponential disk and sky parameters fixed at the best-fit values found in the previous step. We were only able to fit the host galaxy with an exponential disk and flat sky level using this method for IC 342, NGC 3621, and NGC 4244. For the rest of the sample, the scale length and total magnitude were flagged as a problematic parameters by \texttt{GALFIT}, indicating that the host galaxy profile is too flat across the image to be modeled by an exponential disk profile. For M33, NGC 247, NGC 300, NGC 2403, NGC 2976, NGC 4395 and NGC 7793, we used a flat sky level to model the host galaxy light and left its surface brightness as a free parameter when fitting the NC surface brightness profile.

\indent For each fit, we used a $401 \times 401$ pixel ($16'' \times 16''$) fitting region centered on the brightest pixel in the NC. This fitting region size is quite large compared to the typical angular size of NCs in our sample, whose effective radii are all well under $1''$. We experimented with a range of different fitting region sizes to quantify the effect of the fitting region size on the fit results and find that these effects are at the ~10 \% level for the nuclear cluster properties. For a single \sersic\ component fit with a flat sky level, increasing the size of the fitting region will lead to an increase in the brightness of the \sersic\ component, an increase in the effective radius, an increase in the \sersic\ index, and a decrease in the sky level. For example, when we performed a single \sersic\ component fit to the F814W image of NGC 2403 using a $150 \times 150$ pixel fitting region, we obtained a magnitude of 15.48 mag, an effective radius of 10.00 pixels, and a \sersic\ index of 1.51 for the NC component. When we performed the same fit using a  $300 \times 300$ pixel fitting region, we obtained a magnitude of 15.41 mag, an effective radius of 10.82 pixels, and a \sersic\ index of 1.64 for the NC component. In order to be able to compare fit results in different filters, it is important to use a consistent fitting region. In order to optimally sample the host galaxy light, we used the largest fitting region size that could be applied to every image while still excluding bad pixels on the edges of the image introduced during the image combination step of \texttt{AstroDrizzle}. 

\indent We experimented with different functions available in \texttt{GALFIT} to fit to the WFC3 images. Exponential, Gaussian, and de Vaucouleurs profiles are not flexible enough to describe the wide range of inner profile slopes seen in the NCs in our sample. Modified King \citep{king}, Nuker \citep{nuker} and \citet{sersic} profiles all provided good fits to the F814W images of M33, NGC 300, and NGC 7793. These NCs have relatively simple structures and serve as instructive cases for comparison of different surface brightness models. In each case, the \sersic\ profile results in the smallest $\chi^{2}$, although the difference in $\chi^{2}$ between fits performed using \sersic\ and King profiles is always at the 10\% level or less. The Nuker profile has five independent parameters and the King profile has four, while the \sersic\ profile is described by only three independent parameters: the total magnitude, effective radius (half-light radius) and \sersic\ index. We used a \sersic\ profile to model the light profile of the NCs in our sample because the \sersic\ profile provides the best flexibility in describing the structure of different NCs using the smallest number of free parameters. To allow for elliptical shapes we left the axis ratio ($b/a$, where $b$ and $a$ are the lengths of the semiminor and semimajor axes, respectively) and the position angle (PA) of the semimajor axis as free parameters in the fits. 

\begin{figure*}
\centering
\includegraphics[scale=1.3]{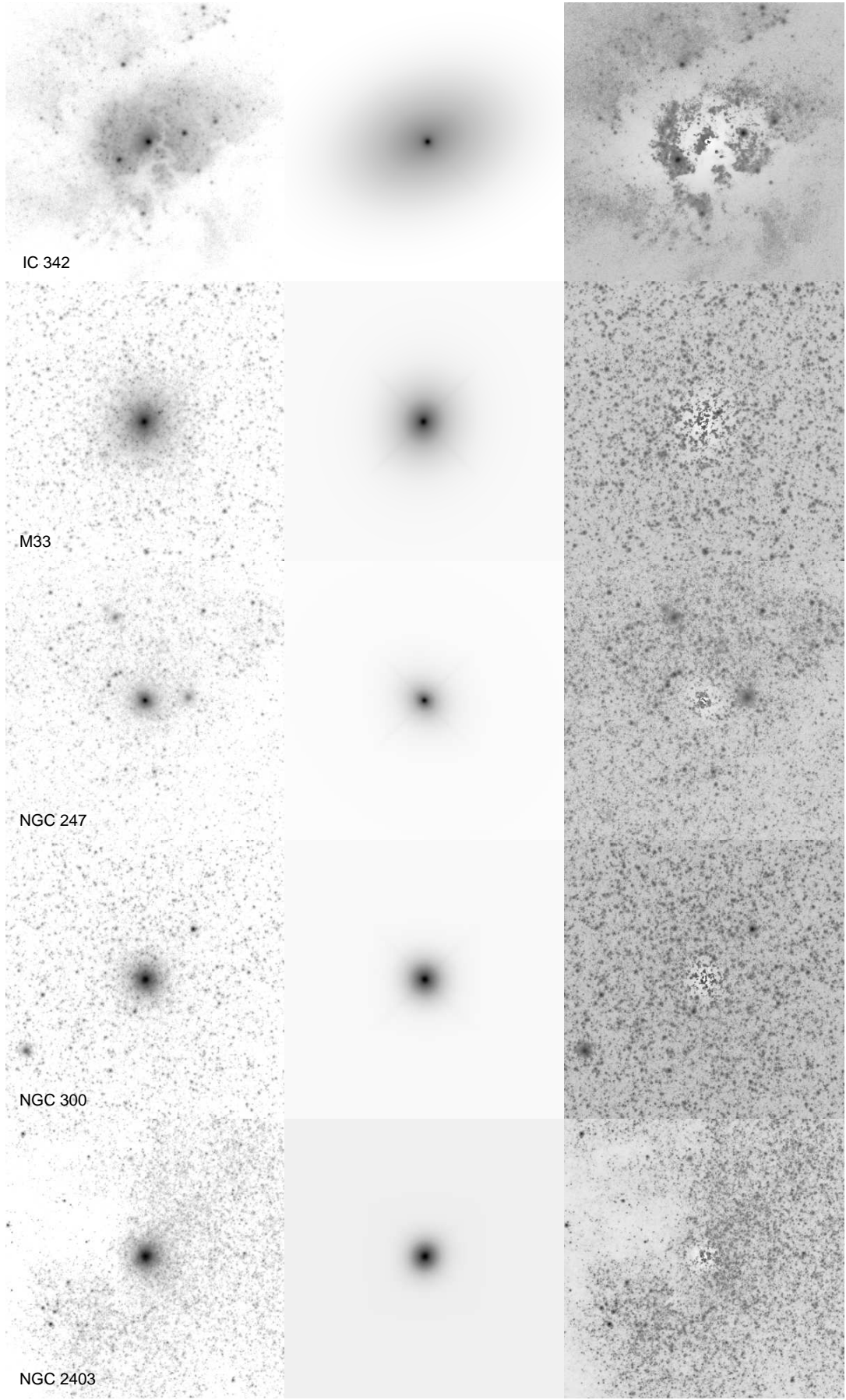}
\caption{Results of the F814W fits. The panels (from left to right) show the F814W image, the best-fit model convolved with the PSF, and the residuals ($image-model$). The field of view is $16'' \times 16''$ for all fits. Images are displayed with their native orientation, as in Figure \ref{all_images}.}
\label{img_fit_res_1}
\end{figure*}

\indent With the exception of NGC 2976, NGC 4395 and NGC 4244, a single \sersic\ profile was used to model each NC in every band. In the bands blueward of F814W, the structure of the NC in NGC 2976 is difficult to fit with any reasonable set of analytic models due to its small-scale clumpiness. This NC also contains a compact blue foreground component to the north (see Figure \ref{master_color_images}). For NGC 2976, we used aperture photometry to measure the total magnitude of the cluster and the magnitude of the blue component in F275W, F336W, and F438W, and F547M. Since the blue component becomes faint and the structure of the NC becomes smoother at longer wavelengths, we used \texttt{GALFIT} to fit a single \sersic\ profile to the cluster in F814W, F127M and F153M. \citet{seth2008} found that the NC in NGC 4244 has two distinct morphological components: a red/old, compact, spheroidal structure, and a blue/young, extended disk structure, while studies by \citet{f&s1989} and \citet{f&h2003} have shown that NGC 4395 hosts a NC as well as an unobscured AGN. Based on these studies, we modeled the NC in NGC 4244 using two \sersic\ components, and for NGC 4395 we used a \sersic\ component to model the NC and a point source component to model the AGN. For comparison to the rest of the sample, we also performed single \sersic\ component fits to the NGC 4244 and NGC 4395 images. A more detailed discussion of individual objects is given in Section 5. Figure \ref{img_fit_res_1} shows the results of the fits to the F814W images for each NC.

\subsection{Comparison with \texttt{ISHAPE}}
\indent \citet{g&b2014} used the \texttt{ISHAPE} procedure in the \texttt{BAOLAB} software package \citep{ishape} to measure effective radii, axis ratios, and magnitudes of a sample of 228 NCs in late-type spiral galaxies using images from the \hst/WFPC2 archive. Much like \texttt{GALFIT}, \texttt{ISHAPE} measures structural properties by minimizing the $\chi^{2}$ difference between an image and a PSF convolved model, but it performs the fit over a circular, rather than a rectangular fitting region. In their study, the NCs were modeled using single component King and power-law models and the \hst/WFPC2 model PSFs were generated using \texttt{TinyTim}. However, they did not allow for the possibility of an exponential disk profile to model the host galaxy light. The substructure seen in some of the clusters in our sample would not be easily detected at the distances typical of NCs in their sample, which was selected using a distance cut of 40 Mpc. As a test, we used \texttt{ISHAPE} to fit single \sersic\ component models to our F814W images using a 200 pixel fitting radius and found very good agreement with our \texttt{GALFIT} results. The mean offset between magnitudes measured using \texttt{ISHAPE} and magnitudes measured using \texttt{GALFIT} (i.e., $ \lvert m_{\mathrm{GALFIT}} - m_{\mathrm{ISHAPE}} \rvert $) was 0.05 mag, while on average, the effective radii and \sersic\ indices agreed within 11\% and 24\%, respectively.

\setcounter{figure}{4}
\begin{figure*}
\centering
\includegraphics[scale=1.3]{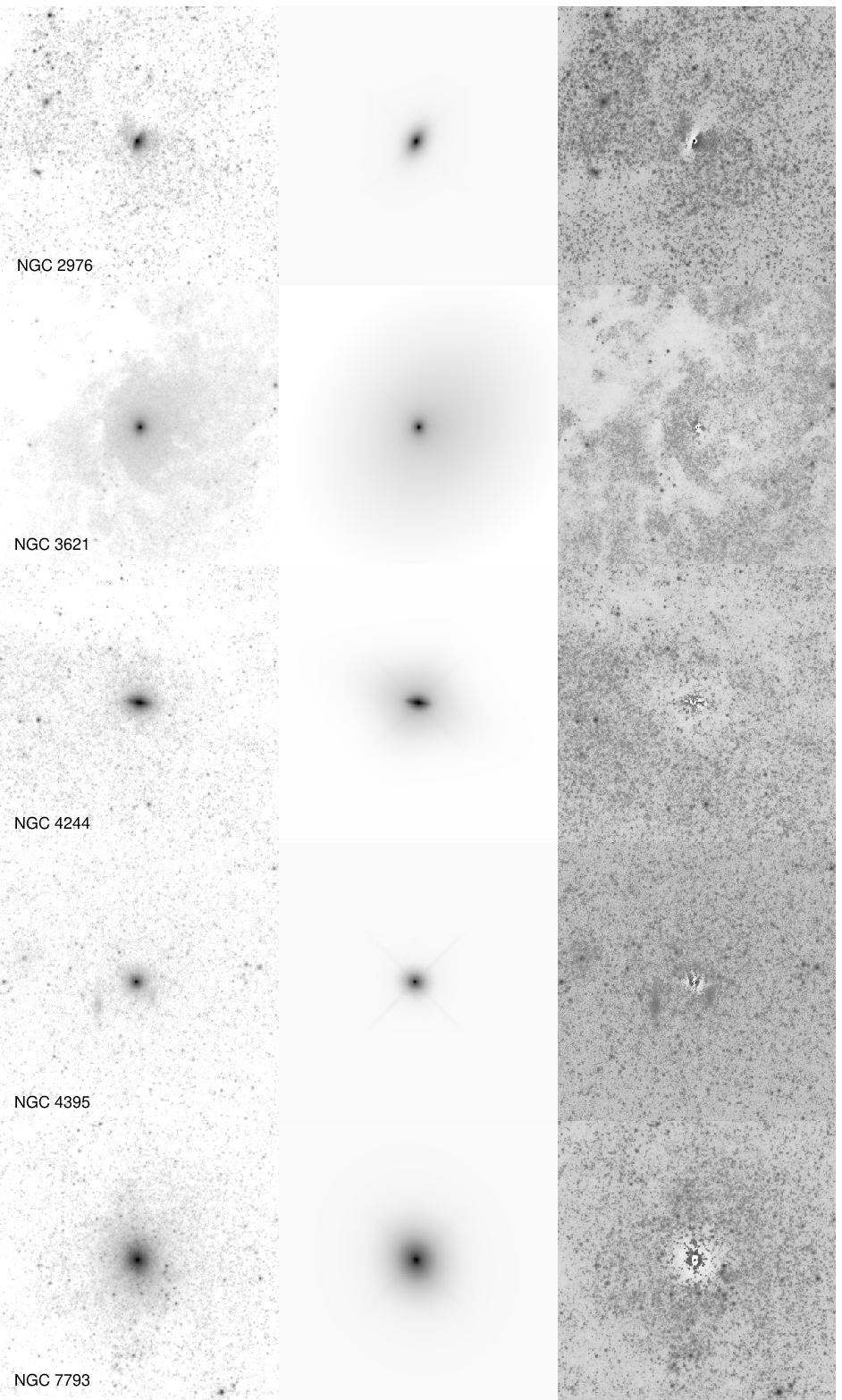}
\caption{Results of the fits in F814W (continued).}
\label{img_fit_res_2}
\end{figure*}

\subsection{1D Radial Profiles}
\begin{figure*} 
\includegraphics[angle=270,scale=0.7]{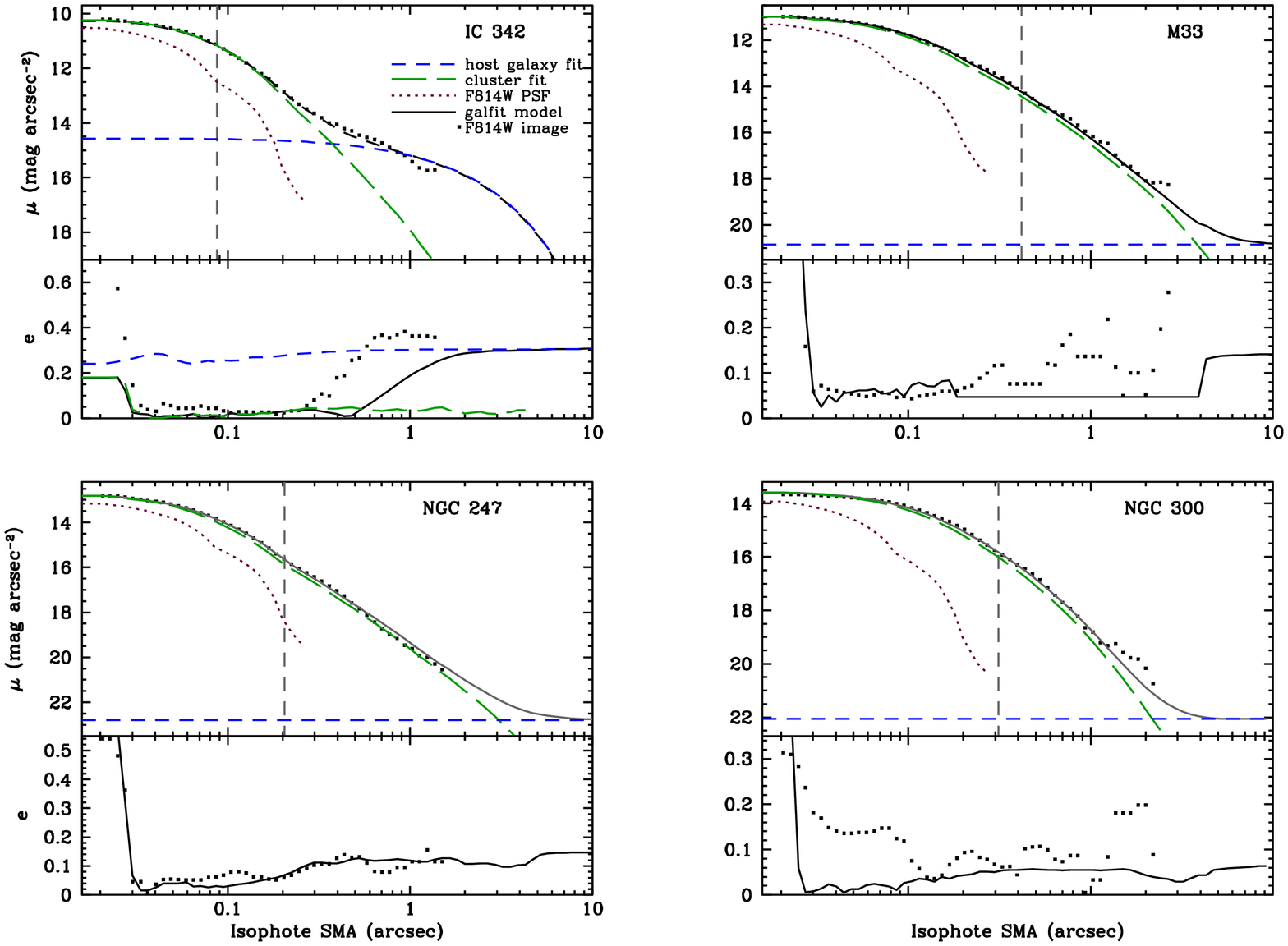}
\caption{F814W radial profiles of the surface brightness and ellipticity of the \texttt{GALFIT} model (solid black line) and \hst/WFC3 image (black squares) as a function of SMA, as well as the subcomponents of the fit. These radial profiles were obtained using the elliptical isophote fitting routine \texttt{Ellipse}. The host galaxy model light profile is denoted by a blue dashed line, and the NC model light profile is denoted by a green long-dashed line. For reference, we include a radial profile of the F814W PSF, indicated by a red dotted line. The vertical gray long-dashed line shows the effective radius of the NC. The convolved NC radial profiles are significantly more extended than the PSF, demonstrating that the clusters are spatially resolved in each case.}
\label{F814W_radial_1}
\end{figure*}

\indent The ``goodness of fit'' is easy to visualize in a 1D radial profile plot, although non-axisymmetric features of the surface brightness profiles are lost in this representation. We fit elliptical isophotes (ellipses of constant surface brightness) to both the \hst/WFC3 images and the PSF-convolved \texttt{GALFIT} models using the PyRAF/STSDAS routine \texttt{Ellipse} \citep{ellipse}, which is based on methods detailed by \citet{ellipse2}. For each isophote, this routine computes the semimajor axis (SMA) of the ellipse, as well as the isophote surface brightness and ellipticity ($e=1-b/a$, where $b/a$ is the axis ratio). We also performed isophotal fits to the model PSFs in order to display their radial surface brightness profiles. Plots of the F814W radial profiles of the surface brightness and ellipticity for each cluster are shown in Figure \ref{F814W_radial_1}. In every case the PSF-convolved \texttt{GALFIT} models are significantly more extended than the PSF and the cluster effective radius exceeds the radius which encloses half of the flux of the F814W PSF (0\farcs063), indicating that all of the NCs in our sample are spatially resolved. 

\setcounter{figure}{5}
\begin{figure*}
%\ContinuedFloat
\centering
\includegraphics[angle=270,scale=0.7]{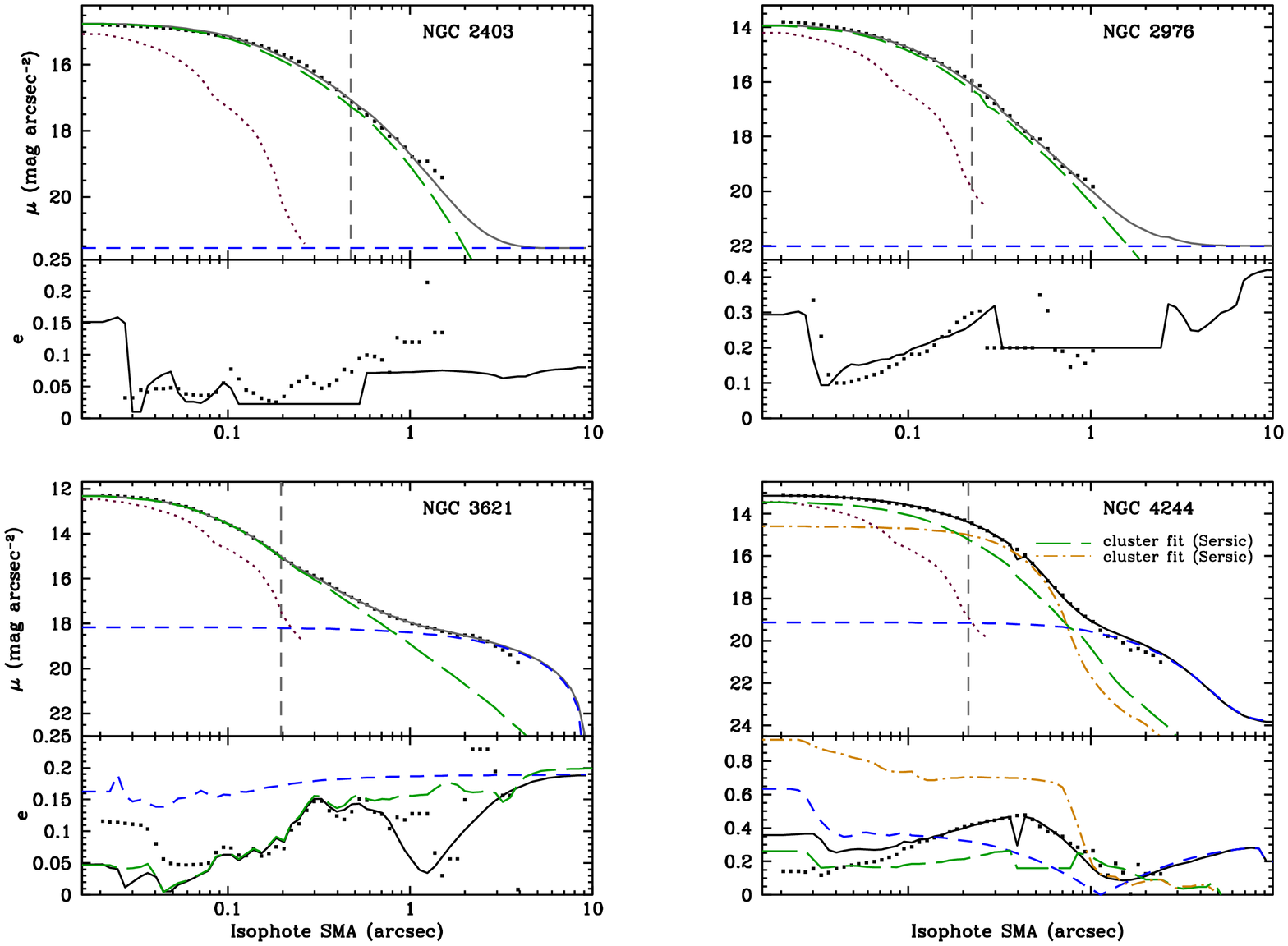}
\caption{F814W radial profile plots (continued). NGC 4244 is fit with 2 \sersic\ components, which are indicated by a green long-dashed line and an orange dash-dotted line.}
\label{F814W_radial_2}
\end{figure*}

\setcounter{figure}{5}
\begin{figure*}
%\ContinuedFloat
\centering
\includegraphics[angle=270,scale=0.7,trim={0cm 0cm 9.5cm 0cm},clip=true]{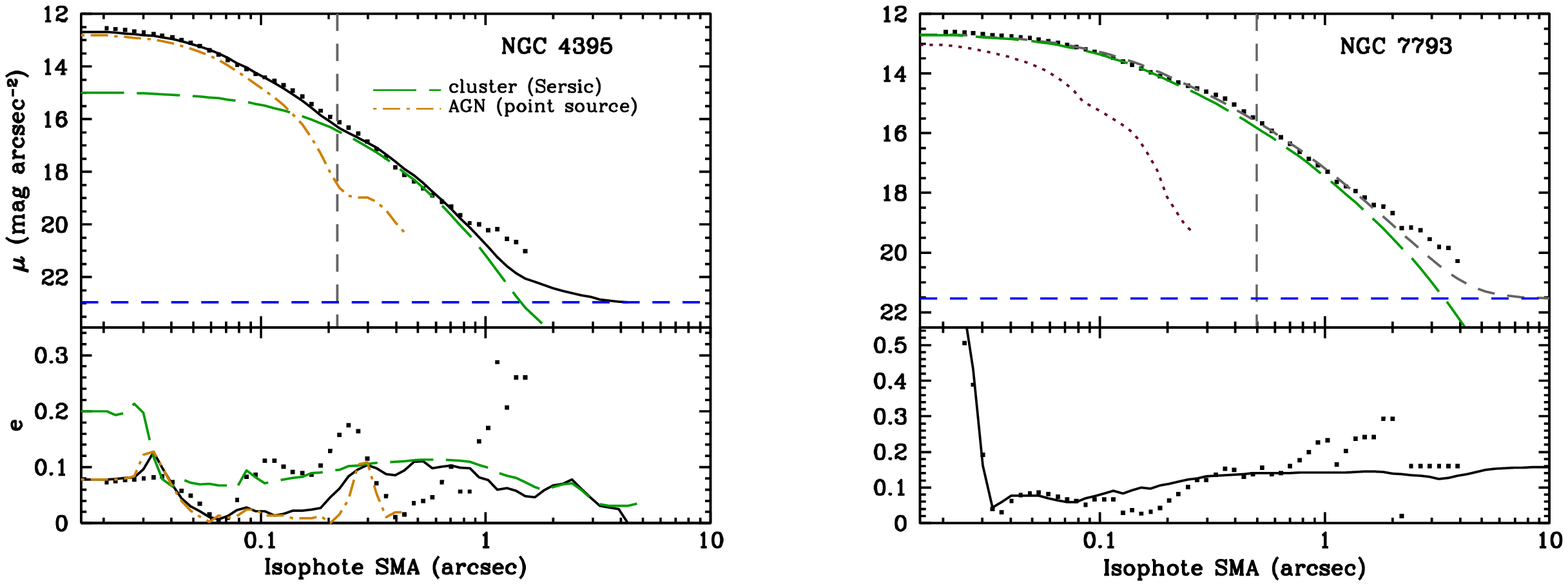}
\caption{F814W radial profile plots (continued). NGC 4395 is fit with a \sersic\ component (indicated by a green long-dashed line) and a point source component (orange dash-dotted line), and the vertical gray long-dashed line denotes the effective radius of the \sersic\ component of this two component fit. We do not include an additional PSF for reference in this plot because the point source is included a fit component.}
\label{F814W_radial_3}
\end{figure*}

\subsection{Uncertainties in Cluster Parameters}
\indent Because the data have very high $S/N$, the formal \texttt{GALFIT} measurement uncertainties in the best-fit parameters obtained from propagating the uncertainty in electron counts at each pixel are quite small. Our error budget is dominated by uncertainties that arise due to the degeneracy between the NC and the background light of the host galaxy light, which we model in \texttt{GALFIT} using either a flat sky level or the combination of a flat sky level and an exponential disk. Estimates of the uncertainties of the best-fit parameters should therefore reflect how sensitive these parameters are to variations in the background level. In general, there is no unambiguous way to define a boundary between the NC and the underlying background light from the disk of the host galaxy. In addition, the background can be quite non-uniform and lumpy, due to the presence of resolved stars, off-nuclear clusters, and patchy dust lanes in the immediate vicinity of the NC. 

\indent The formal uncertainty on the sky parameter returned by \texttt{GALFIT} is typically small ($< 0.1$ ADU), and does not reflect the true uncertainty resulting from the lumpy nature of the immediate surroundings of the NC. In order to more realistically estimate the uncertainty in the sky parameter (\ensuremath{\sigma_\mathrm{sky}}), we took the $401 \times 401$ pixel section of each image that was used in the fit, and divided it into eight equally sized wedges, by splitting it into quadrants along the vertical and horizontal axes of the image, and then splitting each quadrant in half at a $45^{\circ}$ angle with respect to the vertical and horizontal axes. We found the mean sky level in each wedge while masking the NC and computed the standard deviation of these values, to obtain \ensuremath{\sigma_\mathrm{sky}}. This uncertainty (typically a few ADU), is much larger than the formal uncertainty on the sky parameter returned by \texttt{GALFIT}. We used the mean sky level in this calculation, as opposed to the mode sky, because the mean sky is a better representation of the surface brightness of the host galaxy, which includes resolved sources, such as bright stars and off-nuclear clusters. Using the mode sky would underestimate the host galaxy surface brightness by excluding the contribution of individual resolved stars.

\indent For fits where a flat sky level was used to model the background light from the host galaxy, the corresponding uncertainties in the NC best-fit parameters were estimated by re-running \texttt{GALFIT} while holding the sky parameter fixed at \emph{sky} $-$ \ensuremath{\sigma_\mathrm{sky}} (where \emph{sky} is the best-fit sky parameter from the original fit), and again with the sky parameter fixed at \emph{sky} $+$ \ensuremath{\sigma_\mathrm{sky}}. For fits where a combination of a flat sky level and exponential disk were used to model the host galaxy, we re-ran the host galaxy fit as explained in Section 3.2, but held the sky parameter fixed at \emph{sky} $-$ \ensuremath{\sigma_\mathrm{sky}}, and again with the sky parameter fixed at \emph{sky} $+$ \ensuremath{\sigma_\mathrm{sky}}. We then performed the NC fit with the host galaxy exponential disk and sky parameters fixed at the best-fit values found in the previous step to find the upper and lower uncertainties in the NC best-fit parameters. These uncertainties were added in quadrature to the formal measurement uncertainties computed by \texttt{GALFIT} in order to produce the total uncertainty on each fit parameter.

\indent The uncertainty on the distance to each galaxy is an additional source of systematic uncertainty that affects absolute magnitudes in different bands in the same way. In order for the uncertainties that we quote to reflect only our own analysis, we did not include the uncertainty on the distance modulus in the error budget for absolute magnitudes. However, interested readers can refer to the uncertainties on distance moduli quoted in Table \ref{sample}.

\section{Results}
\subsection{Single Band Results}
\subsubsection{Structural Properties}
\begin{figure*}
\includegraphics[scale=0.8,trim={-1cm -1cm -1cm -1cm},clip=true]{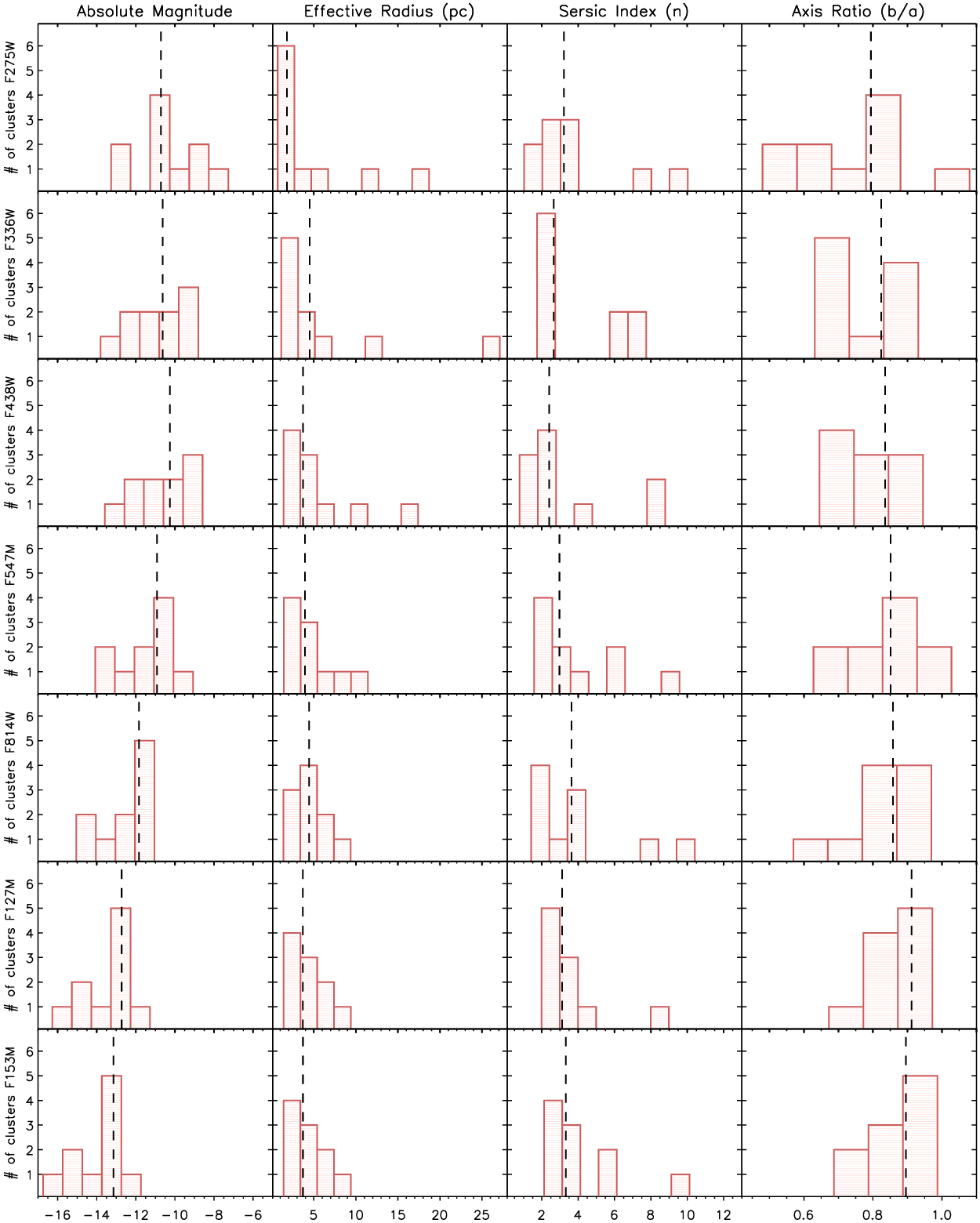}
\centering
\caption{Histograms of the best-fit parameters in each band. For NGC 4244 and NGC 4395, the results of the single \sersic\ component fits are shown. From left to right, each row contains (for a given band) histograms of absolute magnitudes, effective radii, \sersic\ indices, and axis ratios. The median value for each parameter is denoted by a vertical dashed line.}
\label{master_histogram}
\end{figure*}

\indent The NCs in our sample have a wide range of structural properties. From single \sersic\ component fits, we find that F814W absolute magnitudes range from $-11.20$ mag to $-15.05$ mag with a mean of $-12.56$ mag, effective radii range from $1.38$ to $8.28$ pc with a mean of $4.27$ pc, \sersic\ indices range from $1.63$ to $9.73$ with a mean of $3.93$, and axis ratios range from $0.57$ to $0.94$ with a mean of $0.83$. The best-fit parameters in each band are shown in Table \ref{all_params}. We also include the extinction along the line of sight to each galaxy in this table. To help visualize the properties of the sample as a whole, we constructed histograms of the best-fit absolute magnitudes, effective radii, \sersic\ indices, and axis ratios for the single \sersic\ component fits in each band, which are shown in Figure \ref{master_histogram}. 

\begin{figure}
\xput[0.47]{\includegraphics[scale=0.75]{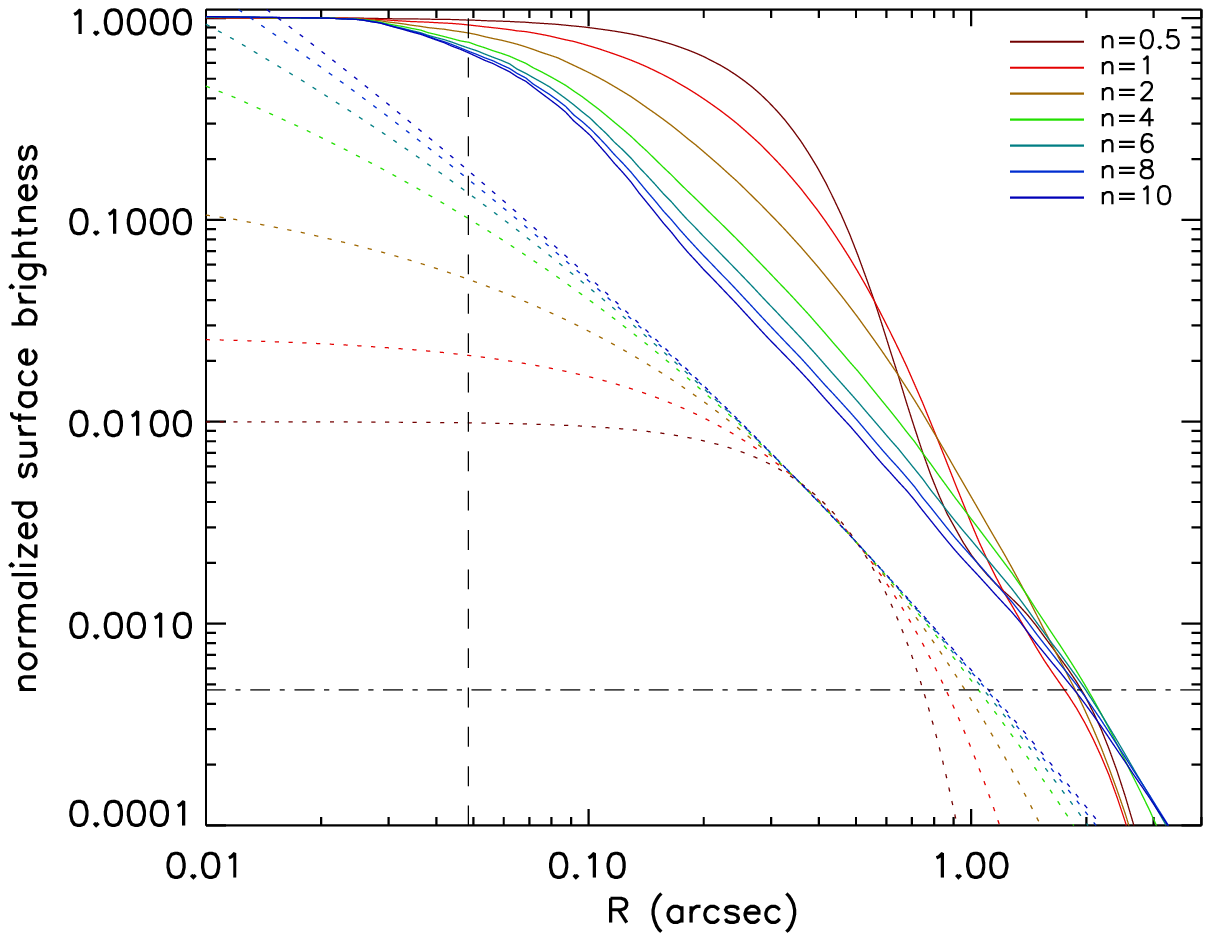}}
\centering
\caption{\sersic\ models for $\Reff = 0\farcs23$, with and without convolution with the F547M PSF. Solid lines denote \sersic\ models that have been convolved with the F547M PSF, while the dotted lines denote intrinsic \sersic\ models which have not been convolved with the PSF. Convolved models are normalized to the same surface brightness at $R = 0$, while unconvolved models are normalized at $R = 0\farcs23$. The black vertical long-dashed line indicates the half width at half maximum of the F547M PSF. The black horizontal dash-dotted line indicates the ratio between the central surface brightness of the host galaxy and the central surface brightness of the NC in F814W, averaged over all ten objects in the sample.}
\label{high_n_test}
\end{figure}

\indent We measure a much wider range of best-fit \sersic\ indices than in previous studies, including some very large values ($n > 8$). For the small number of NCs for which a measurement of the \sersic\ index has been performed, previous studies have generally found $n \sim 1$--$3$ \citep{g&s2009,seth2010}. In order to investigate the robustness of these high concentrations, we used \texttt{GALFIT} to produce noise-free \sersic\ models with an effective radius of 0\farcs23 (which is the mean value for our sample) and \sersic\ indices ranging from $0.5$ to $10$, all which were convolved with the F547M PSF. We show their radial surface brightness profiles, along with radial profiles of the corresponding intrinsic \sersic\ models (which have not been convolved with the PSF) in Figure \ref{high_n_test}. The large values of the \sersic\ index that we measure are determined more by the behavior of the profile in the wings than in the core, because the PSF blurs out information about the slope of the inner profile. For radii inside the half width at half maximum of the PSF, the profiles for $n > 6$ are not distinguishable. However, these profiles are strongly divergent at radii outside of the core of the PSF. Therefore, the large \sersic\ indices that are seen in some objects serve as useful indicators of the shape of the NC profile outside of the core of the PSF. We conclude that NCs have a wide range of concentrations as indicated by the wide range of \sersic\ indices, and that \sersic\ indices above $~6$ do not indicate differences in core concentrations, but instead are dependent on the wings of the profile. 

\begin{deluxetable*}{ccccccccc}
\renewcommand{\arraystretch}{1.5}
\tablecolumns{9}
\tablewidth{0pc}
\tablecaption{NC best-fit parameters}
\tablehead{ &           & F275W                   & F336W                    & F438W                  & F547M                  & F814W                  & F127M                  & F153M      } 
\startdata
IC 342   &\extinction  & $2.94$ & $2.44$ & $2.00$ & $1.51$ & $0.90$ & $0.41$ & $0.30$ \\
         & \appmag      & $17.25^{+0.25}_{-0.08}$ & $16.21^{+0.10}_{-0.23}$ & $15.99^{+0.08}_{-0.17}$ & $15.01^{+0.07}_{-0.14}$ & $13.43^{+0.06}_{-0.08}$ & $11.71^{+0.05}_{-0.10}$ & $11.14^{+0.06}_{-0.10}$ \\
         & \absmag      & $-13.27^{+0.25}_{-0.08}$ & $-13.80^{+0.10}_{-0.23}$ & $-13.59^{+0.08}_{-0.17}$ & $-14.07^{+0.07}_{-0.14}$ & $-15.05^{+0.06}_{-0.08}$ & $-16.27^{+0.05}_{-0.10}$ & $-16.74^{+0.06}_{-0.10}$ \\
         & \reff        & $1.42^{+0.48}_{-0.12}$ & $1.38^{+0.48}_{-0.16}$ & $1.41^{+0.38}_{-0.13}$ & $1.42^{+0.33}_{-0.14}$ & $1.38^{+0.19}_{-0.13}$ & $1.40^{+0.20}_{-0.09}$ & $1.40^{+0.20}_{-0.09}$ \\
         & \sersicn     & $1.03^{+0.63}_{-0.63}$ & $1.94^{+0.95}_{-0.50}$ & $2.56^{+0.88}_{-0.36}$ & $2.97^{+0.78}_{-0.36}$ & $3.79^{+0.48}_{-0.40}$ & $2.47^{+0.73}_{-0.32}$ & $2.49^{+0.69}_{-0.32}$ \\
         & \axisratio   & $0.98^{+0.02}_{-0.01}$ & $0.84^{+0.07}_{-0.07}$ & $0.89^{+0.02}_{-0.01}$ & $0.92^{+0.02}_{-0.01}$ & $0.94^{+0.01}_{-0.01}$ & $0.91^{+0.01}_{-0.01}$ & $0.92^{+0.01}_{-0.01}$ \\
\enddata
\label{all_params}
\tablecomments{Best-fit parameters and their corresponding uncertainties for each NC. We also include the extinction in each band (\ensuremath{A}) along the line of sight to each galaxy. Some axis ratios have very small uncertainties, which round to $0.00$. We list an uncertainty of $0.01$ in these cases. Best-fit parameters for all 10 NCs are available in the online version of the table.}
\end{deluxetable*}

\subsubsection{Size vs.\ Luminosity}
\begin{figure}
\centering
\xput[0.47]{\includegraphics[scale=0.75]{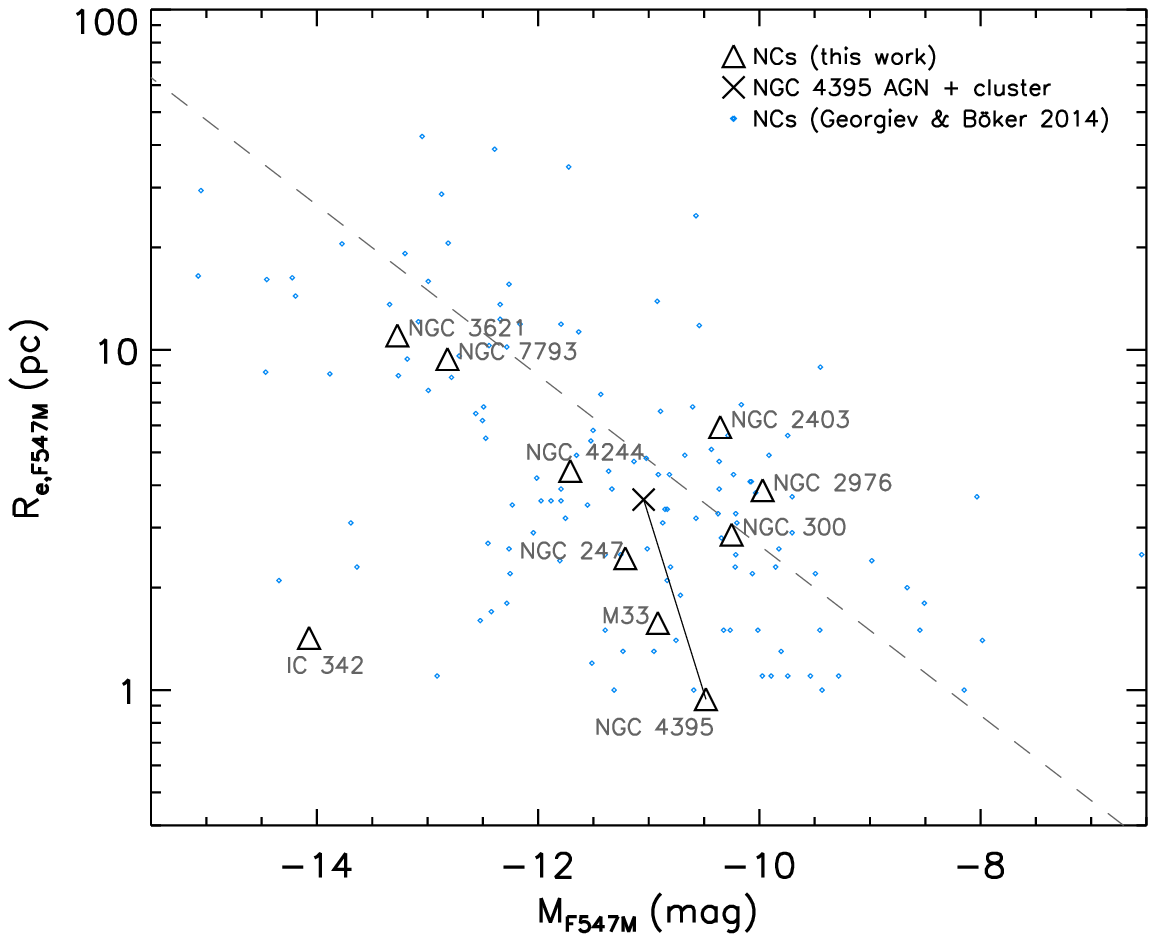}}
\caption{The effective radius measured in the F547M filter vs.\ the F547M absolute magnitude. Black triangles denote NCs from this work. The black ``x'' indicates the single \sersic\ component fit to NGC 4395 (that is, the effective radius and magnitude of the NC plus AGN). The small light blue diamonds indicate NCs from \citet{g&b2014} and the gray dashed line indicates the best-fit \emph{V} band size-luminosity relation derived from their data set.}
\label{size_luminosity}
\end{figure}

\indent We show the F547M effective radius plotted against the F547M absolute magnitude for the NCs in our sample (Figure \ref{size_luminosity}) for comparison with the NCs and the derived size-luminosity relation from \citet{g&b2014}. The magnitudes that they measured in the WFPC2 F606W and F555W filters were converted to \hst/WFC3 F547M magnitudes using the PyRAF/STSDAS task \texttt{Calcphot}, assuming a 5 Gyr, solar metallicity simple stellar population (SSP) for the spectral shape (this is the same spectrum that they used to convert between \hst/WFPC2 magnitudes and Johnson magnitudes). For NGC 2976, we show the F547M absolute magnitude and the F814W effective radius. Most of our clusters fall below the \citet{g&b2014} best-fit size-luminosity relation, which may be in part due to our small sample size. In addition, because they used a large distance cut (40 Mpc), 11\% of the NCs in their sample were not spatially resolved and thus were not used to derive their size-luminosity relation. As a result, the subset of spatially resolved NCs in their sample may be somewhat biased towards larger objects. It is also worth noting that the NC in IC 342 is exceptionally bright and compact. This NC is by far the brightest in our sample, and has a F547M absolute magnitude brighter than 95\% of the NCs in the \citet{g&b2014} sample.  

\subsubsection{NC Magnitude vs.\ Host Galaxy Magnitude}
\begin{figure}
\xput[0.5]{\includegraphics[scale=0.75]{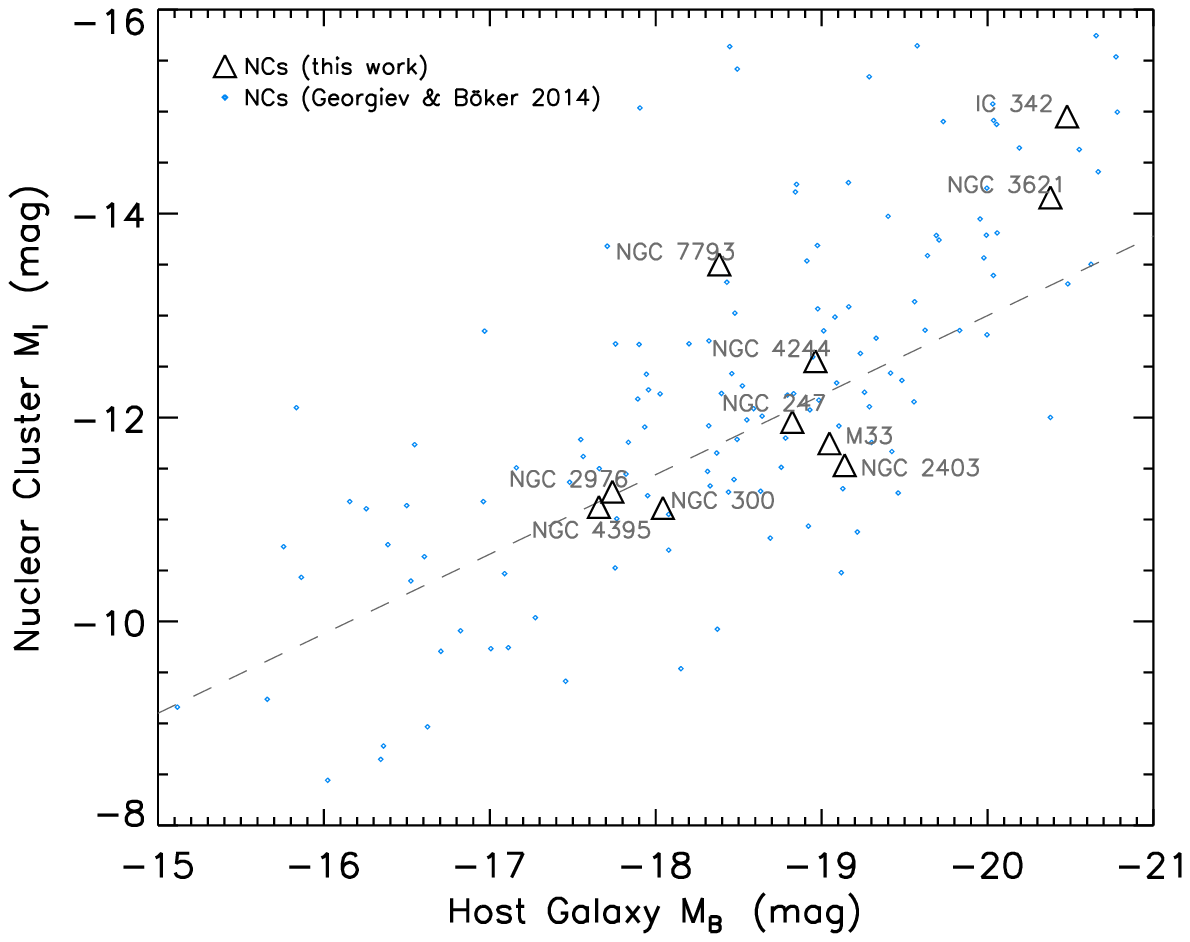}}
\centering
\caption{\emph{I}-band absolute magnitudes of NCs vs.\ \emph{B}-band absolute magnitudes of their host galaxies. Black triangles denote NCs from this work, and small light blue diamonds indicate NCs from \citet{g&b2014}. The gray dashed line indicates the best fit relation from \citet{boker2004}.}
\label{NC_mag_vs_host_mag}
\end{figure}

\citet{boker2004} found that the luminosities of nuclear clusters in late-type spiral galaxies generally correlate with the total luminosities of their host galaxies. In particular, the \emph{I}-band luminosities of NCs show a strong correlation with the \emph{B}-band luminosity of the host, and a weak correlation with the far-IR luminosity of the host, which suggests that NC masses are more governed by the mass of the host galaxy rather than the total star formation rate of the host galaxy. In Figure \ref{NC_mag_vs_host_mag}, we plot the \emph{I}-band absolute magnitudes of the NCs in our sample against the \emph{B}-band absolute magnitudes of their host galaxies, which were taken from NED. We also include the NCs from \citet{g&b2014} and show the best-fit relation from \citet{boker2004}. As described above, \hst\ magnitudes were converted to Johnson magnitudes using \texttt{Calcphot}, assuming a 5 Gyr, solar metallicity SSP for the spectral shape, and extinctions in the Johnson \emph{B} and \emph{I} bands were determined using the method described in Section 2.2. Most of the objects in our sample lie close to the best-fit relation and have luminosities that are quite typical for NCs in late-type spiral galaxies. 

\subsection{Panchromatic Results}
\subsubsection{Color Gradients}
\begin{figure}
\centering
\xput[0.5]{\includegraphics[angle=270,scale=0.36]{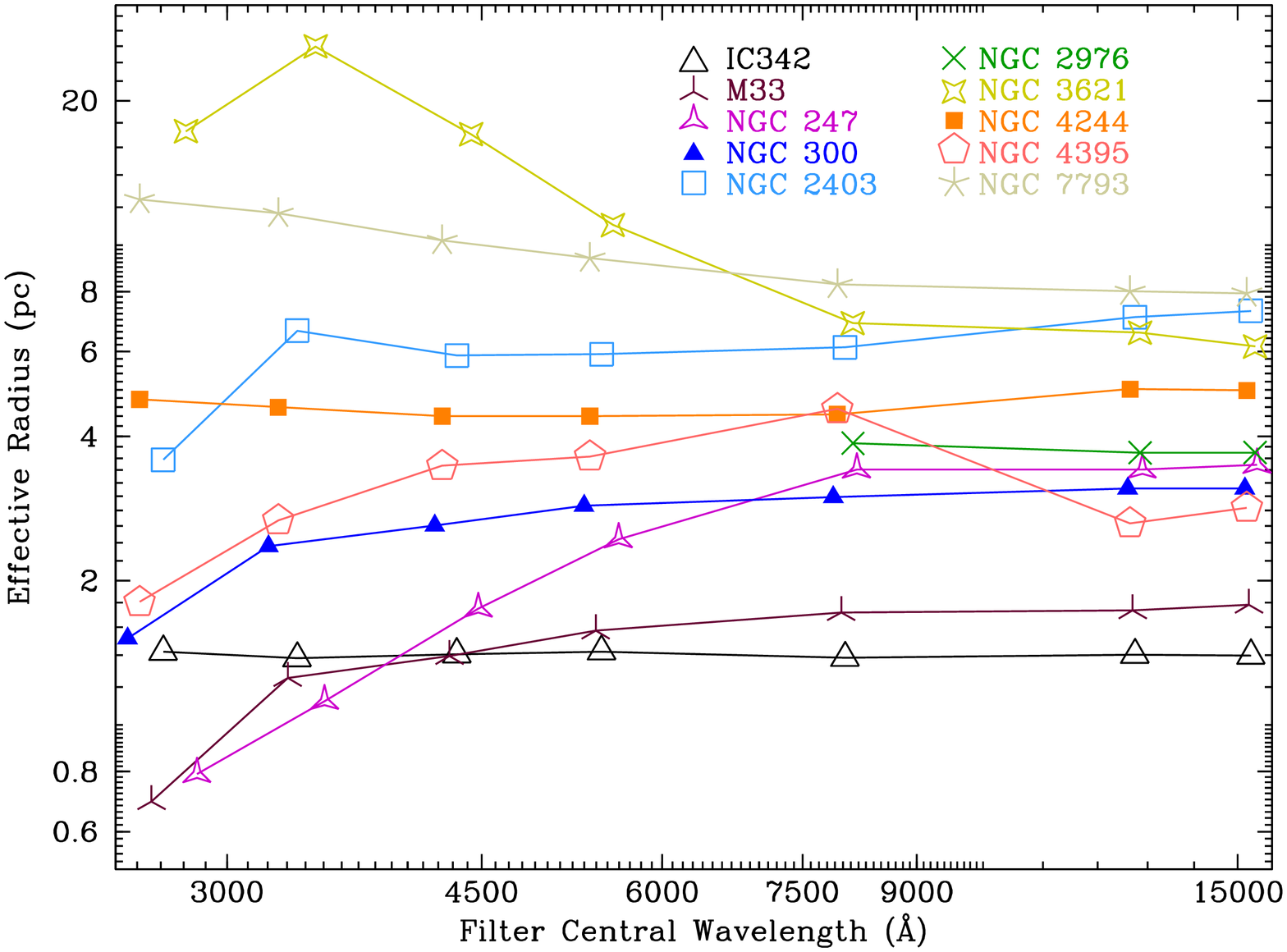}}
\caption{Effective radius of the NCs in each filter. For NGC 4244 and NGC 4395, the results of the single \sersic\ component fits are shown.}
\label{Re_lam}
\end{figure}

\indent We find a strong wavelength dependence of the the effective radius in the majority of the clusters, indicating radial color gradients. In Figure \ref{Re_lam}, we plot the best fit effective radius of each NC against wavelength, showing the results of the single \sersic\ component fits for NGC 4244 and NGC 4395. We performed a formal power-law fit to each curve shown in Figure \ref{Re_lam} to compute their slopes [\Rslope] and corresponding uncertainties. For IC 342, NGC 2976, NGC 2403, and NGC 4244, the best-fit slope is consistent with zero at the $1$ $\sigma$ confidence level. Positive slopes were found for M33, NGC 247, NGC 300, and NGC 4395 (best-fit slopes of $0.24 \pm 0.06$, $1.00 \pm 0.13$, $0.53 \pm 0.02$, and $1.00 \pm 0.01$, respectively). This suggests the presence of a young population that is more concentrated than the bulk of the stars in the NC, or in the case of NGC 4395, indicates the presence of the AGN. These results are consistent with results from the recent \hst/WFPC2 archival study by \citet{g&b2014}, which found that on average, NCs have larger effective radii when measured in redder filters. However, we find a trend of \emph{decreasing} effective radius with wavelength in NGC 3621 and NGC 7793 (best-fit slopes of $0.95 \pm 0.07$ and $-0.30 \pm 0.10$, respectively), which may indicate recent circumnuclear star formation. It should be noted that the larger extent of NGC 3621 in UV bands may be due to UV photons from the central engine of the AGN scattering off narrow-line region clouds rather than a radial gradient in the stellar populations. These color gradients will be explored in more detail in future work, where we will perform spatially resolved stellar population modeling of the NCs in this sample.  

\subsubsection{Increasing Roundness With Wavelength}
\begin{figure}
\centering
\xput[0.47]{\includegraphics[angle=270,scale=0.36]{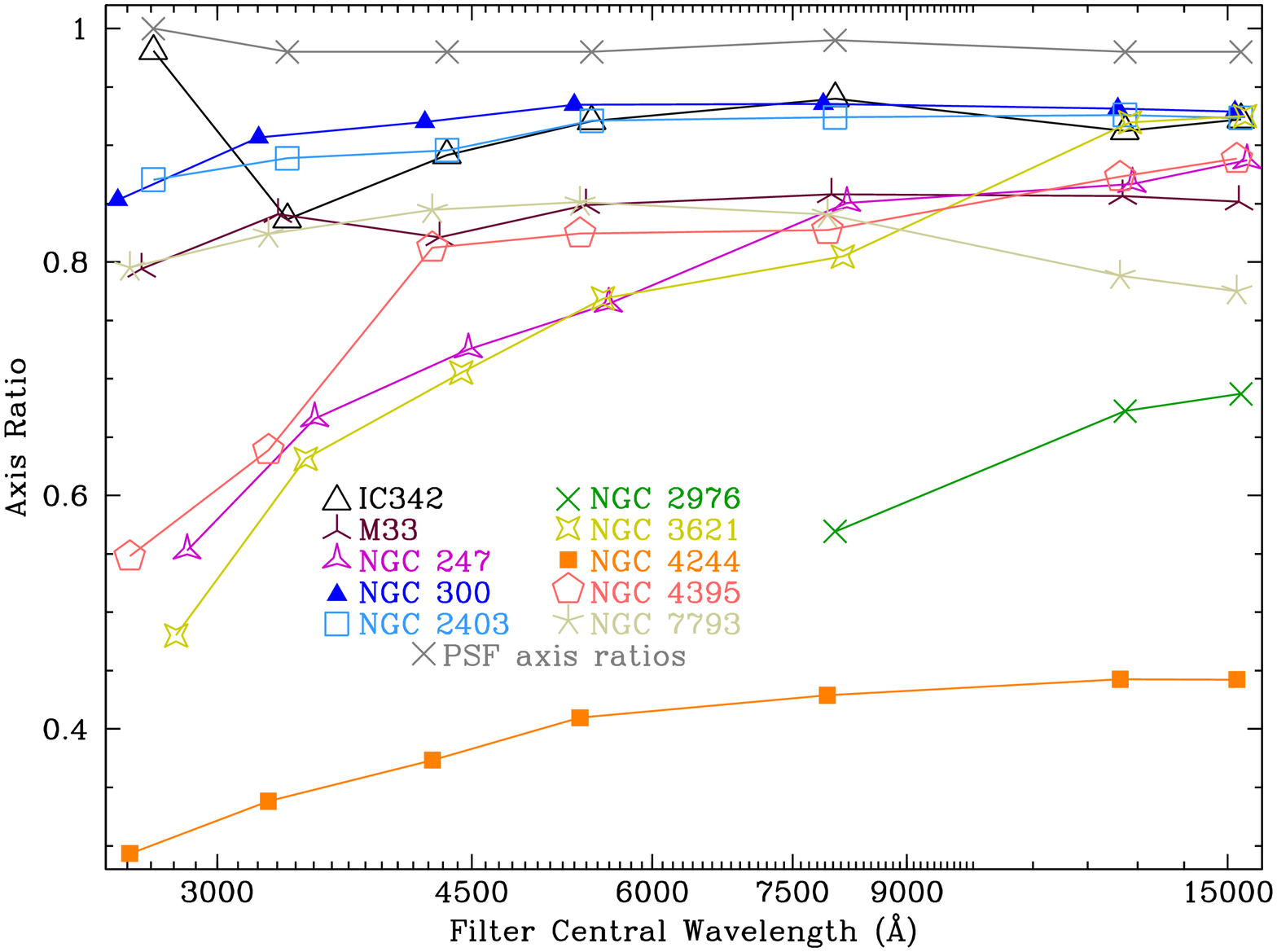}}
\caption{Axis ratios of the NCs in each filter. For NGC 4244 and NGC 4395, the results of the single \sersic\ component fits are shown. The gray x's denote the axis ratio of the PSF in each filter.}
\label{q_lam}
\end{figure}

\indent In a majority of the clusters, we also find a trend of increasing roundness with wavelength. In Figure \ref{q_lam}, we plot best-fit NC axis ratios vs.\ wavelength, showing the results of the single \sersic\ component fits for NGC 4244 and NGC 4395. We also include the axis ratios of the PSF in each band, all of which are nearly unity ($> 0.98$). We performed a formal power-law fit to each curve shown in Figure \ref{q_lam} to obtain best-fit slopes [\qslope] and their corresponding uncertainties. For every NC in the sample, this slope is not consistent with zero at the $1$ $\sigma$ confidence level. IC 342 and NGC 7793 have slightly negative slopes, while every other NC has a positive slope, indicating a monotonically increasing axis ratio with wavelength. This suggests that the young stars in NCs often form in flattened disks, regardless of whether these disks are more extended or compact than the older stellar populations in the cluster. If a cluster contains a young population of stars in a flattened disk structure in addition to an older population in a spheroidal structure, the 2D axis ratio should be closer to unity when measured in redder bands, provided that the disk component is not face-on. On the other hand, if the young and old populations have identical morphologies, the axis ratio should not exhibit any trends with wavelength, but would not necessarily be equal to unity. Thus, if the young stars in NCs typically form in disks, while the older population is in a spheroidal distribution, we would expect objects containing a young stellar disk to exhibit a monotonic increase in axis ratio with wavelength. In NGC 4244, for example, the cluster is clearly being viewed edge-on, and the best-fit axis ratio from the single \sersic\ component fits is $0.29^{+0.01}_{-0.01}$ in F275W and $0.44^{+0.01}_{-0.01}$ in F153M. This is consistent with a previous study by \citet{seth2008}, which showed that the NC contains morphologically distinct old and young populations.  

\indent If the half-light radius of the cluster is significantly smaller than that of the PSF, the intrinsic axis ratio of the cluster will not be recovered in the fit. In the limit of spatially unresolved sources, the best-fit axis ratio will be completely determined by the axis ratio of the PSF, which is always near unity. For our sample, the clusters are sufficiently well resolved that we were able to recover information about their axis ratios. With the exception of IC 342, all the clusters have effective radii larger than the PSF half-light radius in all bands. This suggests the variations in size and flattening with wavelength that we are observing are robust, indicating a difference in the spatial distribution of young and old stellar populations.

\indent The flattening preferentially seen in the bluest bands indicates that the young stars form in disks, possibly fueled by the accretion of low angular momentum gas onto the center of the galaxy. However, simulations by \citet{hartmann2011} demonstrated that the star cluster migratory/merger scenario can also produce NCs with young, blue disks, such as in NGC 4244. These simulations also showed that the star cluster migratory/merger and gas infall formation scenarios predict different kinematics in the resulting NC. Spatially resolved measurements of the kinematics of the NCs are needed to distinguish between these formation scenarios.

\begin{figure}
\centering
\xput[0.45]{\includegraphics[scale=0.8]{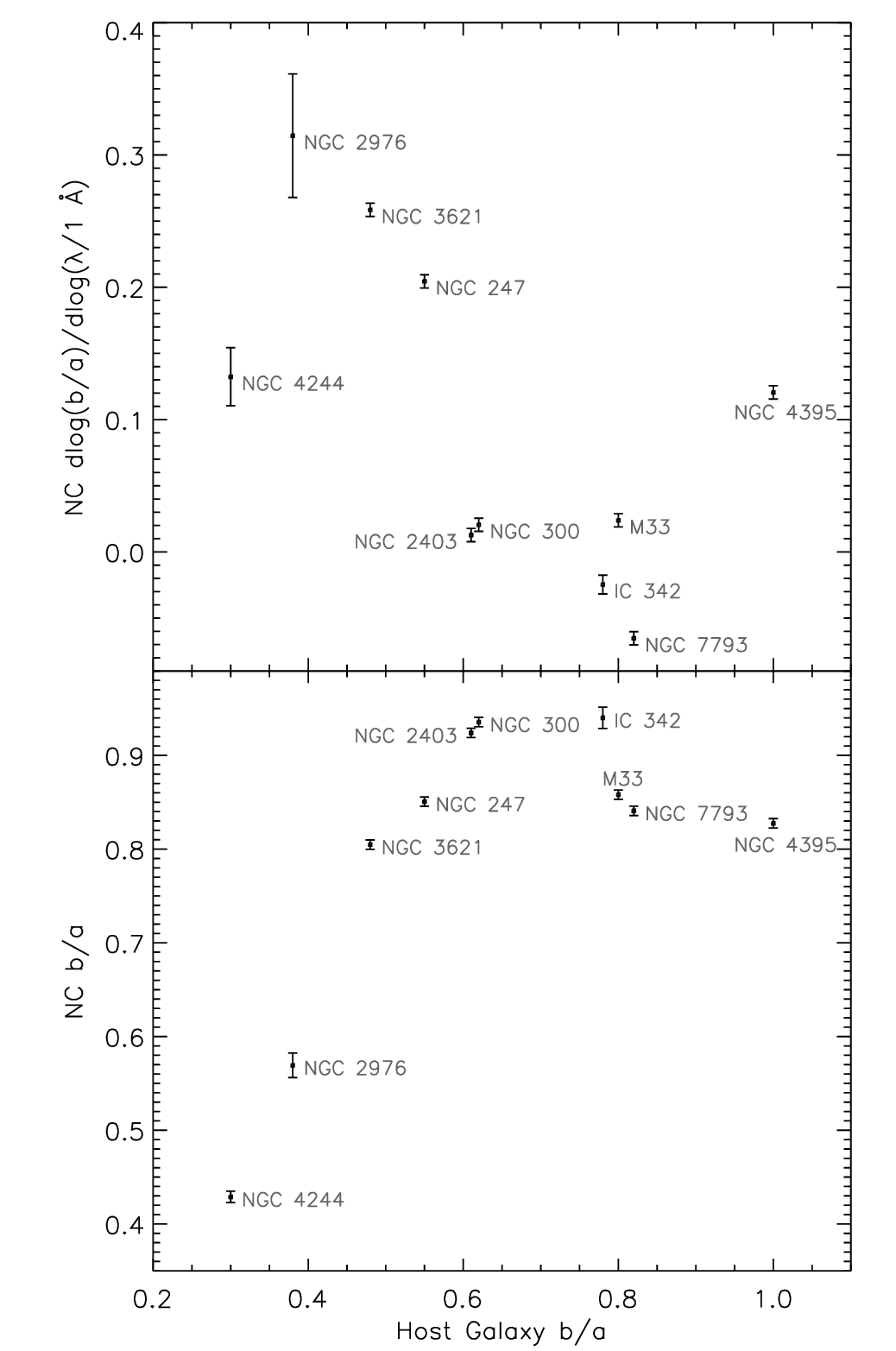}}
\caption{Top panel: \qslope\ of each NC vs.\ the axis ratio of the host galaxy. These slopes [\qslope] were calculated by performing a formal power-law fit to the curves shown in Figure \ref{q_lam}. Bottom panel: axis ratios of the NCs measured in F814W vs.\ the axis ratios of their host galaxies. }
\label{host_vs_NC_q}
\end{figure}

\indent In Figure \ref{host_vs_NC_q}, we plot \qslope\ (top panel) and F814W axis ratios (bottom panel) of the NCs against the axis ratios of the host galaxy disk, which were calculated using 2MASS \emph{K}-band images \citep{2MASS}. The galaxies in our sample with the smallest axis ratios contain the NCs that have the strongest trend of increasing roundness with wavelength and are the most flattened (this can also be seen in Figure \ref{master_color_images}). The \hst/WFC3 images of the NCs in NGC 2976, NGC 3621, and NGC 4244 clearly show the presence of a disk which has the same orientation as the disk of the host galaxy. In each of these cases, the PA of the NC as measured by \texttt{GALFIT} is nearly aligned with the PA of the host galaxy ($\mathrm{\Delta PA} < 20^{\circ}$). Conversely, none of the galaxies in our sample which are close to face-on contain flattened NCs. We conclude that blue disks which lie in the plane of the host galaxy are a common feature of NCs in late-type galaxies, but are difficult to detect in galaxies that are close to face-on. 

\subsection{Stellar Populations}
\subsubsection{Nuclear Cluster SEDs}
\indent For each cluster, we measured a total apparent magnitude using either \texttt{GALFIT} or aperture photometry in all seven bands, allowing us to construct a spectral energy distribution (SED) spanning the near-UV to the near-IR in wavelength. We corrected for Galactic extinction using dust maps from \citet{s&f2011}, as explained in Section 2. Magnitudes were converted to flux units (erg s\per\ cm\persq\  \AA\per) using \texttt{Calcphot}, then finally converted to luminosities using the distances given in Table \ref{sample}. The resulting SEDs are shown in Figure \ref{SEDs}. Color-color diagrams (see below) provide a way to visualize the comparison between the observed SEDs and model spectra. Fits of multi-component stellar population models to the NC SEDs are needed to more precisely characterize the stellar populations present in each NC, and will be described in a future paper. 

\begin{figure}
\centering
\xput[0.51]{\includegraphics[angle=270,scale=0.36]{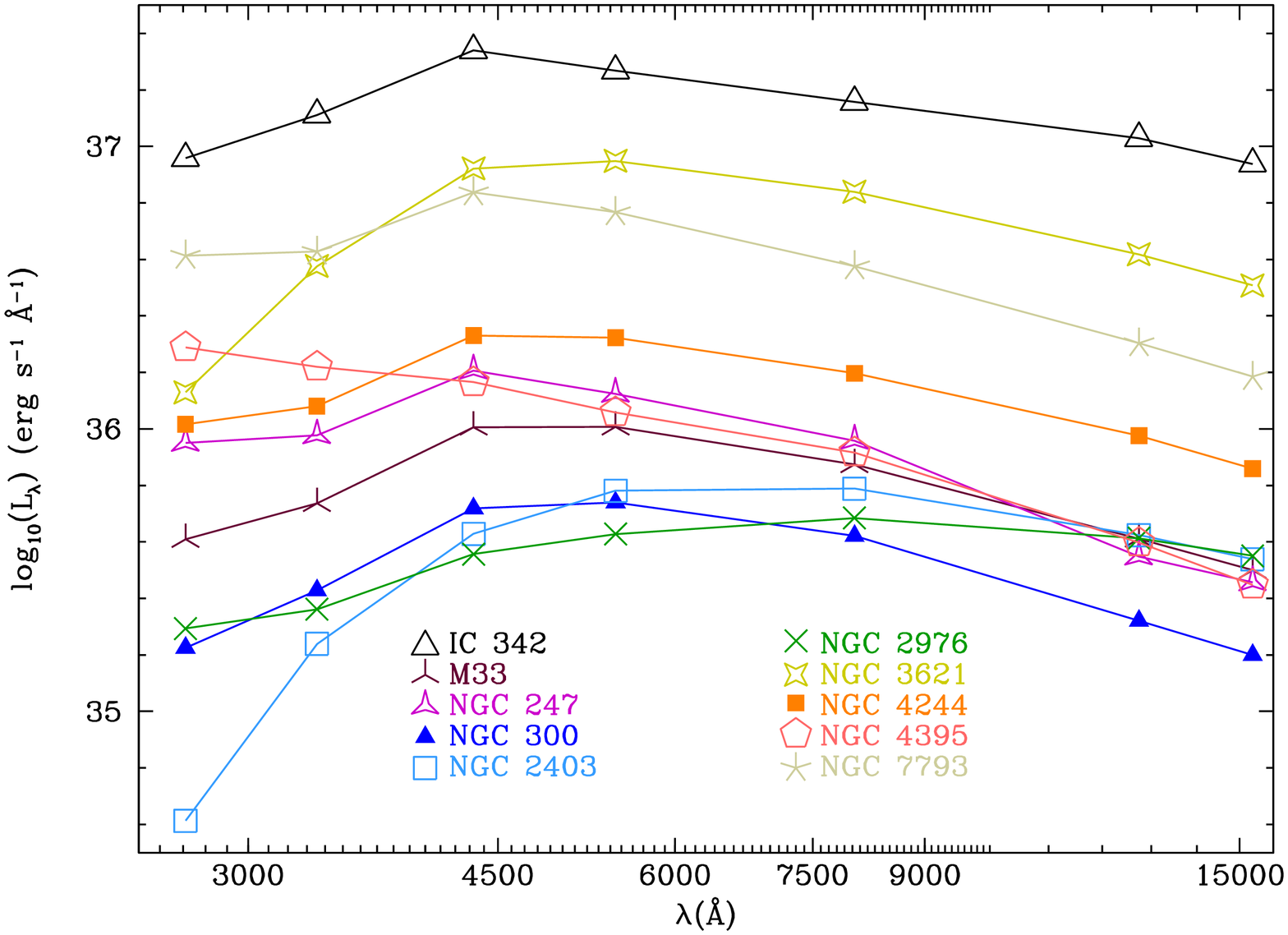}}
\caption{SEDs of the 10 NCs in our sample. For NGC 4244 and NGC 4395, we show the composite SED of the two components of the fit.}
\label{SEDs}
\end{figure}

\subsubsection{Nuclear Cluster Colors}
\begin{figure} 
\centering
\xput[0.47]{\includegraphics[scale=0.65]{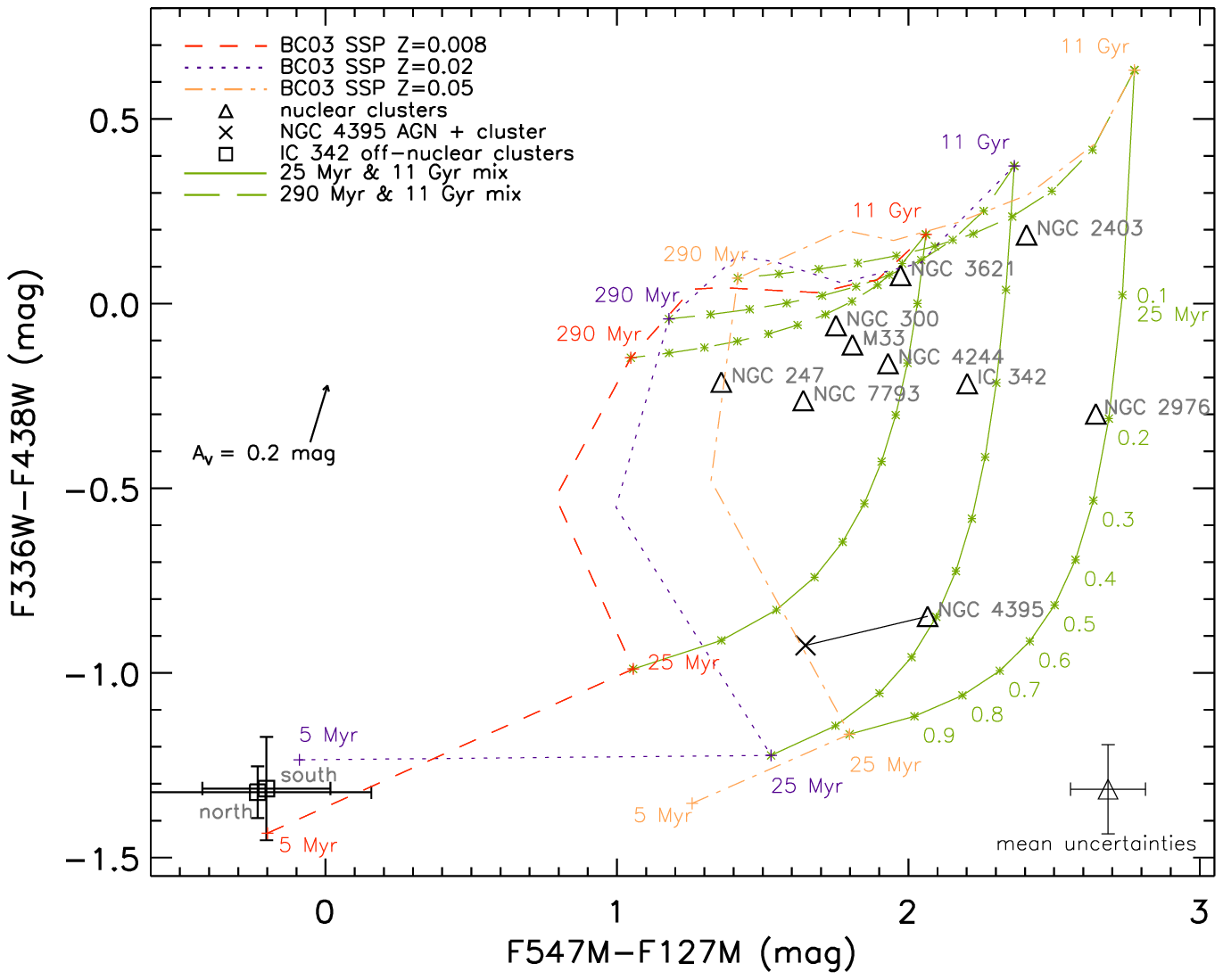}}
\caption{F336W $-$ F438W vs.\ F547M $-$ F127M colors. Black triangles denote NCs from this work, and the off-nuclear clusters in IC 342 are shown as black squares. For NGC 4395, we show colors of the NC as well as the combined colors of the NC and AGN (denoted by a black ``x''). We include evolutionary tracks of SSPs from \citet{bc2003} with $\emph{Z} = 0.008$ (denoted by the red dashed line), $\emph{Z} = 0.02$ (purple short-dashed line), and $\emph{Z} = 0.05$ (orange dash-dot line). Mixing curves for a combination of \citet{bc2003} SSPs with age 25 Myr and 11 Gyr are denoted by the solid gray green lines, and mixing curves for a combination of 290 Myr and 11 Gyr SSPs are denoted by the long-dashed gray green lines. We include mixing curves for metallicity $\emph{Z} = 0.008$, $0.02$, and $0.05$. The arrow indicates a reddening vector of $A_{V} = 0.2$ mag. The error bars in the lower right corner show the mean uncertainty in the colors on each axis (averaged over all NCs in the sample).}
\label{color_color_2}
\end{figure}

\begin{figure} 
\centering
\xput[0.5]{\includegraphics[scale=0.65]{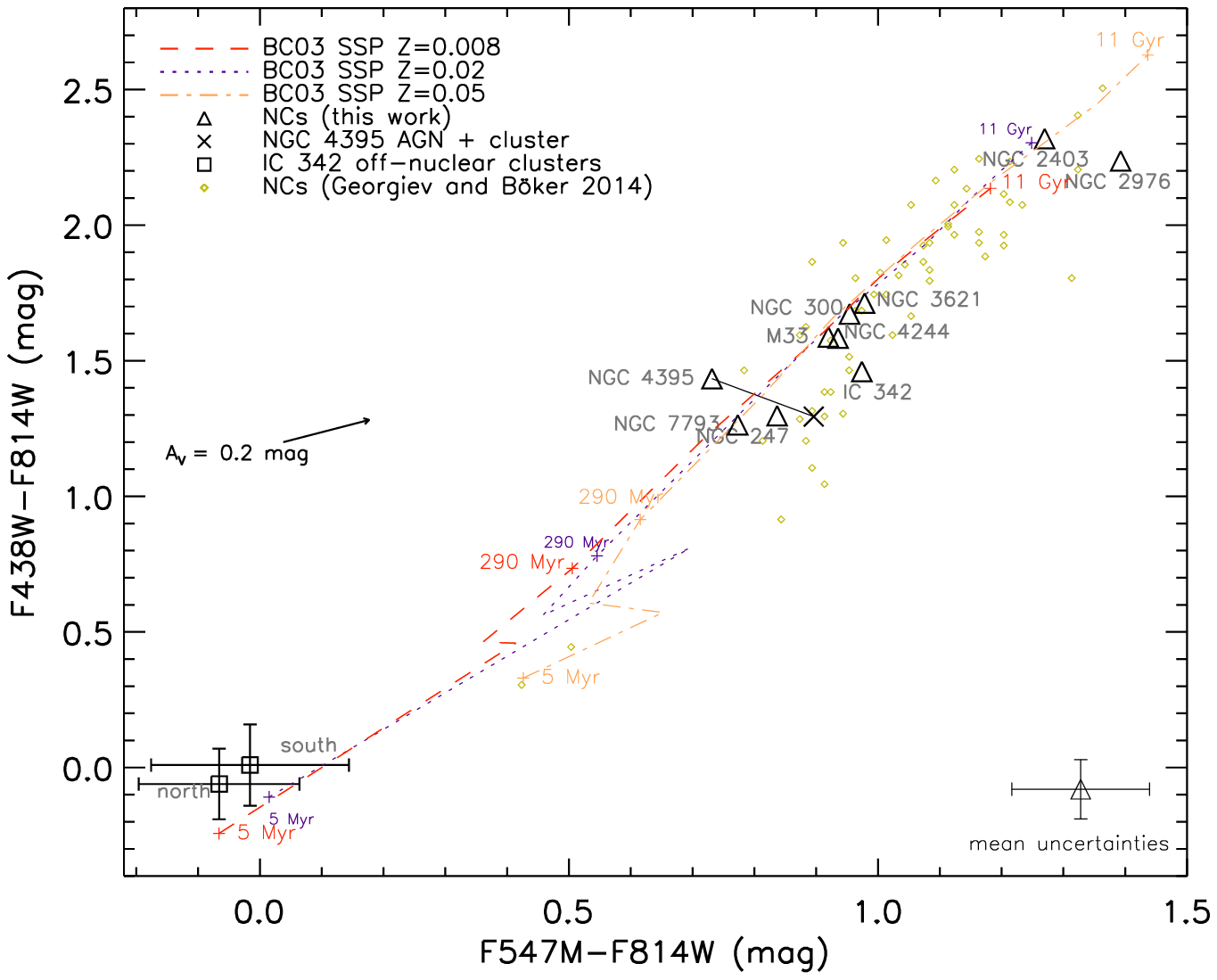}}
\caption{F438W $-$ F814W vs.\ F547M $-$ F814W colors. We include NC colors from \citet{g&b2014}, denoted by small gold diamonds. We did not include mixing curves in this plot because the model tracks are nearly straight lines in this color-color space, especially at older ages.}
\label{color_color_3}
\end{figure}

\indent In order to characterize their stellar populations, we computed NC colors from our extinction-corrected absolute magnitudes and compared them to colors of single-age simple stellar populations from \citet{bc2003} for metallicity $\emph{Z} = 0.008$, $0.02$, and $0.05$. We created two color-color diagrams, one spanning UV to the IR, and one restricted to the optical bands, which are given in Figures \ref{color_color_2} and \ref{color_color_3}, respectively. For NGC 4395, we show colors of the NC (\sersic\ component, denoted by a black triangle) as well as the combined colors of the NC and AGN (\sersic\ plus point source component, denoted by a black ``x''). We also include colors of the NCs from \citet{g&b2014} on the optical only color-color plot. The magnitudes that they measured in the WFPC2 F450W, F606W, and F814W filters were converted to magnitudes in the \hst/WFC3 F438W, F547M, and F814W filters using \texttt{Calcphot}, assuming a 5 Gyr, solar metallicity SSP for the spectral shape. In this plot, our NCs fall close to the SSP tracks and are broadly consistent with the colors of the NCs from \citet{g&b2014}. 

\indent  Most of the NCs in our sample lie at locations distinctly offset from the SSP tracks in F336W $-$ F438W vs.\ F547M $-$ F127M color-color space, confirming that a mix of stellar populations with different ages is present in many of the NCs. This is expected, given that previous studies have shown that the star formation histories of NCs are generally not well described by a single starburst event \citep{walcher2006,rossa2006}. In order to investigate the mix of ages, we took linear combinations of \citet{bc2003} SSP spectra with ages 11 Gyr and 25 Myr and computed their colors in order to construct a ``mixing curve'', which traces the color of the total population as the fractional contribution from the 11 Gyr population to the total luminosity is adjusted from $1.0$ to $0.0$ and the contribution from the 25 Myr population to the total luminosity is adjusted from $0.0$ to $1.0$, in steps of $0.1$. A mixing curve for a combination of a 11 Gyr and 290 Myr SSP is also included in the UV-optical-IR color-color plot. The typical age mix for the NCs in our sample is a combination of an old ($> 1.4$ Gyr) and a young ($100$ - $300$ Myr) population. The mix of ages is in the NCs is obvious when considering colors that span the UV to the IR, but is not apparent from the optical colors only. These color-color plots reinforce the conclusions of \citet{demeulenaer2014}, who highlighted the importance of broad wavelength coverage in stellar population studies, and demonstrated how degeneracies between age, metallicity and extinction, which limit stellar population studies that utilize optical bands only, can be broken with full UV to IR coverage. 

\section{Comments on Individual Objects}
\subsection{IC 342}
\indent IC 342 is a face-on giant Scd spiral galaxy located in the Maffei Group at a distance of 3.3 Mpc \citep{saha2002}. In a dynamical study of IC 342 using CO bandhead spectra and \hst/WFPC2 images, \citet{boker1999} measured an integrated stellar velocity dispersion of 33 \kms\ for the NC, estimated the total stellar mass to be $6 \times 10^{6}$ \msun, and placed an upper limit on the mass of any central BH of $5 \times 10^{5}$ \msun. Due to its proximity to the Galactic plane, there is significant foreground extinction, particularly in the UV bands (\ensuremath{A_\mathrm{F275W}}$=2.94$ mag). There are dust lanes to the east and west of the main cluster that we masked out during the fits. We were able to fit a model consisting of an exponential disk and a flat sky level to the host galaxy in every band, and the nuclear cluster was fit using a single \sersic\ component in every band. \citet{boker1999} measured an apparent \emph{V}-band magnitude of 15.19 mag for the cluster, which corresponds to an extinction-corrected F547M absolute magnitude of $-13.91$ mag, which is somewhat discrepant with our measurement of $-14.07 ^{+0.07}_{-0.14}$ mag. However, the \hst/WFC2 images that they used were saturated in the core of the NC, which may explain the discrepancy. The NC in IC 342 is one of the most compact and by far the most luminous in our sample, with an \emph{I}-band absolute magnitude \ensuremath{M_\mathrm{F814W}}$=-15.05 ^{+0.06}_{-0.08}$ mag. The effective radius is nearly constant with wavelength, with a best-fit value of $1.38^{+0.19}_{-0.13}$ pc in F814W. 

\indent There are several off-nuclear clusters in the central regions of IC 342. The brightest two are located 2\farcs11 (projected distance $33.55$ pc) to the north and 1\farcs92 ($30.53$ pc) to the south of the NC, respectively. Although other galaxies in our sample contain off-nuclear clusters, these two were included in the color-color diagrams in Section 4 because they are by far the most luminous in our sample, and actually outshine the main NC in UV bands. We measured their magnitudes using aperture photometry over a circular aperture (and applied an aperture correction to account for flux lost outside the aperture due to PSF blurring), because \texttt{GALFIT} could not converge on a good fit without the use of constraints. These off-nuclear clusters have extremely blue colors, consistent with 5 Myr SSPs, as shown on the color-color diagrams shown in Figures \ref{color_color_2} and \ref{color_color_3}. We will carry out a more detailed examination, including an analysis of stellar population ages and stellar masses, of the complete set of off-nuclear clusters in our sample in future work.

\subsection{M33}
\indent Located at a distance of 0.9 Mpc \citep{bono2010}, M33 is the only member of the Local Group in our sample. \citet{gebhardt2001} measured an integrated velocity dispersion of 20 \kms\ for the cluster, and placed a very tight upper limit on the mass of any central BH of 1500 \msun. Using the same \hst\ data, \citet{merritt2001} placed an upper limit of 3000 \msun\ on the mass of the central BH. \citet{kormendy2010} estimated the total stellar mass in the cluster to be $10^{6}$ \msun. The nucleus of M33 coincides with the ultraluminous X-ray source X-8 \citep{trinchieri1988}, which is believed to be a stellar-mass BH accreting at a super-Eddington rate \citep{foschini2004}. 

\indent In all bands, we fit a flat sky level to the host galaxy and a single \sersic\ component to the nuclear cluster. The NC in M33 is one of the most compact in the sample. We find a steady increase in the effective radius with filter central wavelength from $0.69^{+0.29}_{-0.10}$ pc in F275W to $1.78^{+0.20}_{-0.13}$ pc in F153M, which indicates that the center of the cluster is bluer than the wings. This is consistent with a \hst/WFPC2 imaging study by \citet{lauer1998}, which found a decrease in the \emph{V}$-$\emph{I} color towards the center of the cluster, and a study by \citet{kormendy1993} which found a similar gradient in the \emph{B}$-$\emph{R} color. 

\subsection{NGC 247}
\indent NGC 247 is a member of the Sculptor Group, at a distance 3.4 Mpc \citep{bono2010}. We fit a flat sky level to the host galaxy and a single \sersic\ component to the NC in all bands. We find evidence for a color gradient within this cluster as well. The effective radius is smallest in the UV and increases towards the IR, with a best-fit value of $0.79^{+0.07}_{-0.07}$ in F275W and $3.49^{+1.80}_{-0.69}$ in F153M. The best-fit axis ratio is $0.55^{+0.01}_{-0.01}$ in F275W and monotonically increases to $0.89^{+0.01}_{-0.01}$ in F153M, which suggests that the young stars formed in a disk.

\subsection{NGC 300}
\indent At a distance of 2.0 Mpc \citep{bono2010}, NGC 300 lies in the field between the Local Group and the Sculptor Group. \citet{walcher2005} measured a stellar velocity dispersion of 13.3 \kms\ and a mass of $10^{6.02}$ \msun\ for the NC. We fit a flat sky level to the host galaxy and a single \sersic\ component to the NC in all bands. NGC 300 was included in the \citet{boker2002} \hst/WFPC2 NC imaging survey, where an \emph{I}-band absolute magnitude of $-11.43 \pm 0.40$ mag was measured for the NC. Correcting for the slightly different distances used, this corresponds to a WFC3 F814W absolute magnitude of $-11.08 \pm 0.40$ mag, which is consistent with our measurement of $-11.20^{+0.02}_{-0.02}$ mag. The best-fit effective radius increases monotonically from $1.52^{+0.03}_{-0.03}$ pc in F275W to $3.12^{+0.25}_{-0.16}$ pc in F153M, indicating a color gradient within the cluster. The \sersic\ index is nearly constant across all filter bands, with a value of $2.27^{+0.10}_{-0.08}$ in F814W. 

\subsection{NGC 2403}
\indent NGC 2403 resides in the outskirts of the M81 Group at a distance of 3.1 Mpc \citep{saha2006}. In addition to the NC, NGC 2403 hosts a highly variable central X-ray source, likely to be an X-ray binary system with a stellar mass BH accretor \citep{yukita2007}. We fit a flat sky level to the host galaxy and a single \sersic\ component to the NC in every band. Dust lanes southwest of the NC were masked out during the fits. With an effective radius of 6.13 pc and an absolute magnitude of $-11.62^{+0.11}_{-0.10}$ in F814W, the cluster has a low surface brightness, leading to relatively large uncertainties in the fit parameters. The effective radius of the cluster is generally larger when observed in redder filters, with a value of in $3.57^{+5.08}_{-0.90}$ pc F275W and $7.29^{+3.28}_{-1.23}$ pc in F153M.

\subsection{NGC 2976}
\indent NGC 2976 is located at a distance of 3.6 Mpc \citep{jacobs2009} in the M81 group. In F814W, F153M, and F153M, we were able to fit a single \sersic\ component to the NC while fitting a flat sky level to the host galaxy. In these bands, the cluster is highly flattened and the PA is nearly aligned with that of the host galaxy disk ($b/a = 0.57$ and $\mathrm{\Delta PA} = 13^{\circ}$ in F814W). In the bands blueward of F814W, the structure of the NC is difficult to fit with any reasonable set of analytic models due to its small-scale clumpiness. In these bands, the cluster can be resolved into two distinct structures: a compact blue component to the north and a more extended, irregularly shaped red component to the south (see Figure \ref{master_color_images}). The contrast between these components diminishes at larger wavelengths, and they completely blend together in the F814W, F127M, and F153M images (as can be seen in Figure \ref{all_images}). In order to obtain the total cluster magnitude in F275W, F336W, F438W, and F547M, we performed aperture photometry using an elliptical aperture centered on the centroid of the cluster. We used the axis ratio and position angle of the best-fit \sersic\ model in F814W to define the position angle and axis ratio of the elliptical aperture. The semimajor axis of the aperture we used was $9 \times R_{\mathrm{e,F814W}}$ (where $R_{\mathrm{e,F814W}}$ is the best-fit effective radius for the cluster in F814W), while the inner and outer radii of the sky annulus were $9 \times R_{\mathrm{e,F814W}}$ and $10 \times R_{\mathrm{e,F814W}}$, respectively. These aperture and sky annulus sizes were found to give the best agreement with the total cluster magnitude obtained from \texttt{GALFIT} in F814W, F127M, and F153M. We calculated a mean difference of 0.10 mag between the total cluster magnitude obtained from \texttt{GALFIT} and from aperture photometry in F814W, F127M and F153M, and used this to define the measurement uncertainty. From aperture photometry, we measured the following extinction-corrected absolute magnitudes for the cluster: $M_{\mathrm{F275W}}=-9.10$ mag, $M_{\mathrm{F336W}}=-9.43$ mag, $M_{\mathrm{F438W}}=-9.13$ mag, $M_{\mathrm{F547M}}=-9.97$ mag. 

\indent We also measured the magnitude of the blue component in F275W, F336W, F438W, and F547M by performing aperture photometry over a circular aperture. In each band, we performed the measurement 12 times using aperture radii in the range 2--5 pixels, inner sky radii in the range 5--20 pixels and outer sky radii in the range 10--30 pixels, applying the appropriate aperture correction each time. We calculated the median between the largest and smallest magnitude obtained from the 12 trials, and used this range to define the associated measurement uncertainty. We obtained the following absolute magnitudes for the blue component: $M_{\mathrm{F275W}}=-8.49 \pm 0.09$ mag, $M_{\mathrm{F336W}}=-8.38 \pm 0.22$ mag, $M_{\mathrm{F438W}}=-7.35 \pm 0.50$ mag, $M_{\mathrm{F547M}}=-7.98 \pm 0.78$ mag. For comparison, the brightest Wolf-Rayet and O stars in the R136 cluster in the LMC have \emph{V}-band absolute magnitudes as bright as -7.33 mag \citep{crowther2010}. Because of the large uncertainties in our measurements, we cannot rule out the possibility of the blue component being an extremely luminous, hot star, although it may also be a very young, compact star cluster. We are unable to determine whether this object is associated with the NC or being viewed in projection. 

\subsection{NGC 3621}
\indent NGC 3621 is the most distant galaxy in our sample, at 7.3 Mpc \citep{saha2006}. The NC has a velocity dispersion of 43 \kms, one of the largest ever measured for any NC in a late-type spiral galaxy, which sets a fairly high upper limit on the mass of any central BH of $3 \times 10^{6}$ \msun\ \citep{barth2009}. \citet{satyapal2007} discovered a faint AGN in the NC from mid-IR [\ion{Ne}{5}] line emission and based on estimates of its bolometric luminosity, set a lower limit on the mass of the central BH of $4\times10^{3}$ \msun. \citet{gliozzi2009} found further evidence for an AGN in NGC 3621 with the detection of a weak X-ray point source coincident with the nucleus.

\indent We successfully fit an exponential profile plus a flat sky level to the host galaxy in every band, and we fit a single \sersic\ component to the NC in all seven bands.  Although NGC 3621 contains an AGN, adding a point source component did not improve the quality of the fits, consistent with expectations for a highly obscured AGN. Dust lanes northeast, southeast, and northwest of the NC were masked out during the fits. \citet{barth2009} measured a \emph{J}-band apparent magnitude of 14.14 mag for the NC from archival NICMOS images, which corresponds to a F127M absolute magnitude of $-15.24$ mag, in very good agreement with our measurement of $-15.25^{+0.07}_{-0.13}$ mag. The NC has a very large \sersic\ index ($n>7$) in all bands, with a best-fit value as high as $9.73^{+1.00}_{-1.67}$ in F127M. In the UV, the cluster is highly elongated along the north-south axis, with an axis ratio ($b/a$) of $0.48^{+0.01}_{-0.01}$. This elongated UV emission can also be seen in the color composite image shown in Figure \ref{master_color_images}. The axis ratio monotonically increases with wavelength to $0.93^{+0.01}_{-0.01}$ in F153M. The cluster is also more extended in the UV, with an effective radius of $17.27^{+3.83}_{-3.83}$ pc in F275W and in $6.15^{+0.72}_{-0.72}$ F127M. The greater north-south elongation and larger extent of the cluster in the UV could potentially be explained by UV emission from the central engine being obscured by an edge-on torus oriented along the east-west axis and scattering off narrow-line region clouds to the north and the south. If so, the extended UV emission would be linearly polarized. Alternatively, the extended UV emission could be explained by extended star formation in the outskirts of the cluster. In F275W, the PA of the semimajor axis of the NC is nearly aligned ($\mathrm{\Delta PA} = 19^{\circ}$) with the PA of the host galaxy measured from the 2MASS \emph{K}-band image \citep{2MASS}, which is expected if the NC primarily grows through gas accretion rather than star cluster mergers \citep{seth2006,hartmann2011}. However, there is no correlation between the orientation of AGNs and their host galaxies \citep{fischer2013}, so we cannot determine the nature of the extended UV emission from our UV images alone. UV spectroscopy and/or polarimetry can help to distinguish between young stars and scattered AGN emission. 

\subsection{NGC 4244}
\begin{figure*}
\centering
\includegraphics[angle=270,scale=0.65]{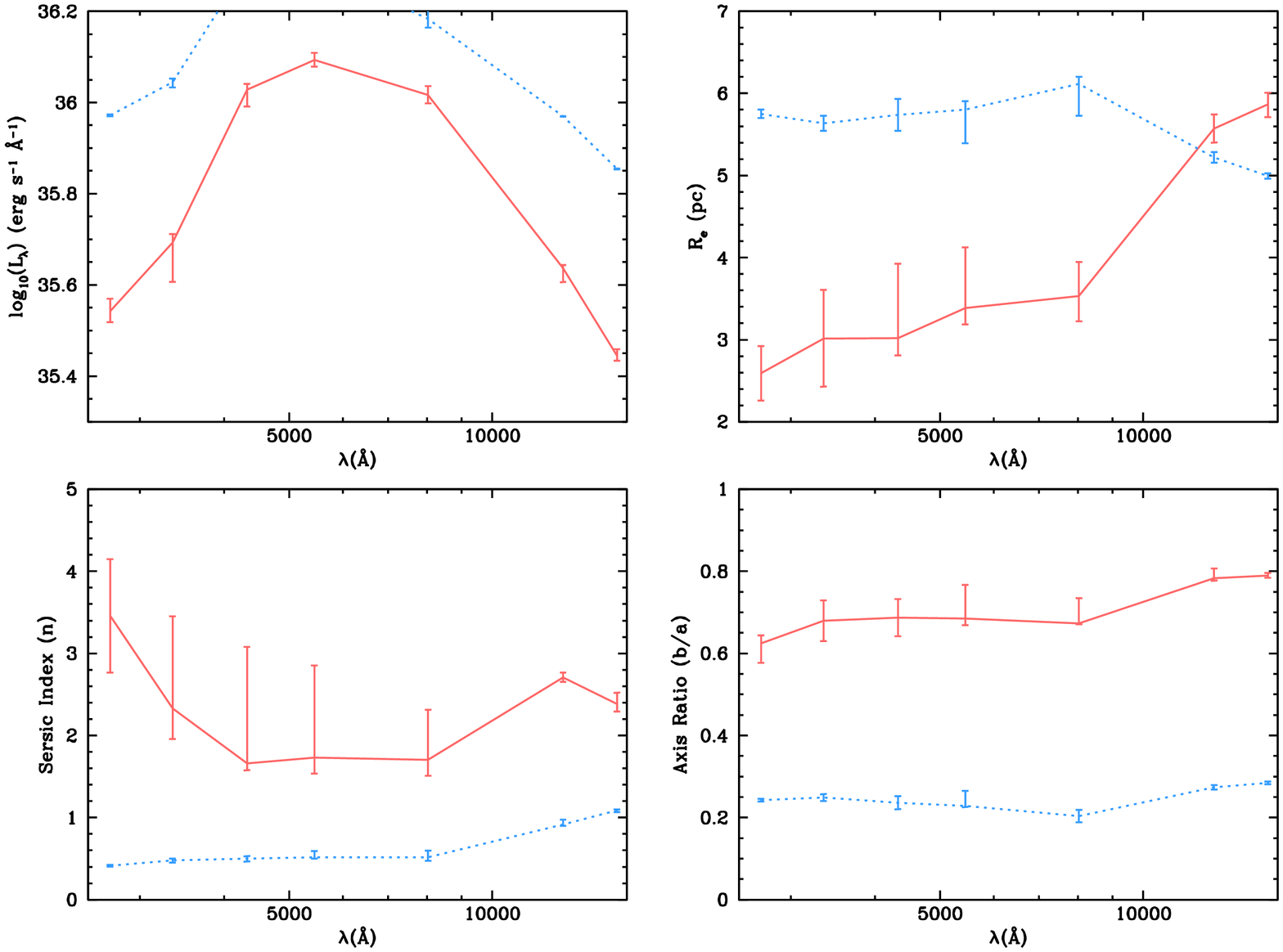}
\caption{Luminosities, effective radii, \sersic\ indices and axis ratios of the individual \sersic\ components as a function of filter central wavelength for NGC 4244. The cluster consists of an extended, highly flattened, blue component and a compact, spheroidal, red component. These components are denoted with a blue short-dashed line and a red solid line, respectively.}
\label{NGC4244_plots}
\end{figure*}

\indent Located at a distance of 4.3 Mpc \citep{jacobs2009} in the M94 Group, NGC 4244 is the only galaxy in our sample with an edge-on orientation. In a recent Gemini Near-Infrared Integral Field Spectrograph (NIFS) study of this NC, \citet{seth2008} discovered rotation of 30 \kms\ in the central 10 pc of the cluster and found that the NC contains two distinct populations: an old, compact spheroidal component and a young, extended disk component. In each band, we fit an edge-on exponential disk plus a flat sky level to the host galaxy, while the NC was modeled with two \sersic\ components. In F814W, the flattened component has an effective radius of $6.12^{+0.08}_{-0.39}$ pc, a \sersic\ index of $0.52^{+0.08}_{-0.04}$, and an axis ratio of $0.20^{+0.02}_{-0.02}$, while the rounder component has an effective radius of $3.53^{+0.42}_{-0.31}$ pc, a \sersic\ index of $1.71^{+0.61}_{-0.20}$, and an axis ratio of $0.67^{+0.06}_{-0.01}$. SEDs of the two components, as well as plots of their effective radii, \sersic\ indices, and axis ratios as a function of wavelength are shown in Figure \ref{NGC4244_plots}. We also fit a single \sersic\ component to the NC in each band for comparison with other clusters. The single \sersic\ component effective radius is nearly constant with wavelength, with a best-fit value of $4.77^{+0.02}_{-0.02}$ pc in F275W and $4.98^{+0.01}_{-0.01}$ in F153M. From the single \sersic\ component fits, we find that the axis ratio monotonically increases from $0.29^{+0.01}_{-0.01}$ in F275W to $0.44^{+0.01}_{-0.01}$ in F153M, and the semimajor axis is nearly aligned with that of the host galaxy in every band ($\mathrm{\Delta PA} = 12^{\circ}$ -- $13^{\circ}$). The alignment of the NC and host galaxy semimajor axes is consistent with previous studies of NGC 4244 \citep{seth2006,hartmann2011,delorenzi2013}, and suggests that gas accretion has played a significant role in the formation of the NC. 

\subsection{NGC 4395}
\indent NGC 4395 is oriented face-on at a distance of 4.3 Mpc \citep{thim2004} in the M94 Group. The existence of a NC in NGC 4395 was explicitly demonstrated by \citet{f&h2003}, from \texttt{GALFIT} analysis of a \hst/WFPC2 \emph{I}-band image. The discovery of faint broad hydrogen and helium lines in its optical spectrum revealed that NGC 4395 contains one of the least luminous Type 1 AGNs known \citep{f&s1989}. A \hst/STIS reverberation-mapping program by \citet{peterson2005} yielded an estimate of $3.6 \times 10^5$ \msun\ for the mass of the central BH, which implies that the AGN is very underluminous and is accreting at just $1.2 \times 10^{-3}$ times the Eddington rate. In a recent photometric reverberation mapping campaign, \citet{edri2012} measured an \hal\ lag of just 3.6 hours, corresponding to a BH mass of $5 \times 10^{4}$ \msun\ (the large discrepancy in the BH mass estimates from this study and Peterson et al. is in part due to the different assumed values of the scaling factor in the BH mass calculation, which depends on the unknown geometry and kinematics of the broad-line region). \citet{minezaki2006} detected \emph{J} and \emph{H} band flux variations over a single night of monitoring of the AGN, in addition to large flux variations in optical to near-IR bands on timescales of days to months.

\begin{figure}
\centering
\xput[0.45]{\includegraphics[scale=0.7]{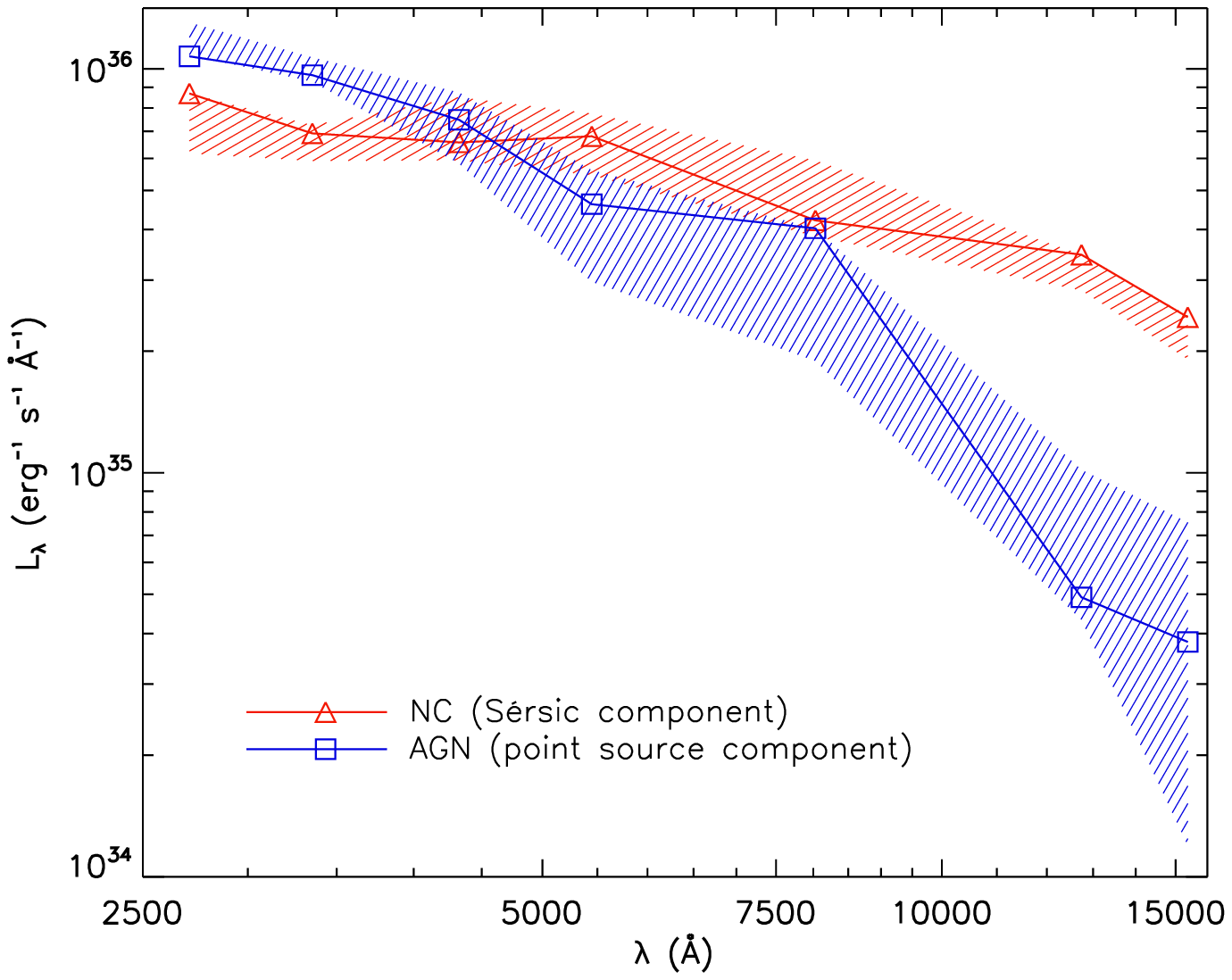}}
\caption{SEDs of the individual fit components for NGC 4395. A point source component (shown in blue) was used to model the AGN and a \sersic\ component (shown in red) was used to model the NC. The solid lines correspond to the SEDs resulting from fits performed with no constraints. In addition to the fit with no constraints, fits were performed with the effective radius fixed at 4.56 pc and the \sersic\ index fixed at 1.42, and with the effective radius fixed at 1.80 pc and the \sersic\ index fixed at 3.01. The shaded regions are bounded by the SEDs resulting from taking the minimum and maximum magnitude of all three fits in each band.}
\label{NGC4395_plot}
\end{figure}

\indent We used a flat sky level to model the host galaxy, a single \sersic\ component for the NC, and included a point source component for the AGN, since the central engine is unobscured along the line of sight. There is an emission line region west of the cluster which was masked during the fits. With no constraints on any of the fit parameters, we find that the \sersic\ component has the largest effective radius in the F814W band (\Reff\ $= 4.56$ pc, $n = 1.42$), and the smallest effective radius in F275W (\Reff\ $= 1.80$ pc, $n = 3.01$). We found no obvious trends in the effective radius or the \sersic\ index with wavelength. In order to examine the degeneracy between the magnitude of the \sersic\ component and the point source component in our fits, we repeated the fits in all seven bands with the effective radius of the \sersic\ component fixed at 1.80 pc and the \sersic\ index fixed at 3.01, and then again with with the effective radius fixed at 4.56 pc and the \sersic\ index fixed at 1.42. The range of magnitudes obtained from these fits gives a rough estimate of the photometric uncertainties due to the degeneracy between the profiles of the point source component and the \sersic\ component in the fits. Figure \ref{NGC4395_plot} shows the SED of both the \sersic\ component and point source component obtained from these fits. From single \sersic\ component fits (with no PSF component), we find that the axis ratio monotonically increases from $0.55^{+0.01}_{-0.01}$ in F275W to $0.89^{+0.01}_{-0.01}$ in F153M.

\indent \citet{f&h&s1993} measured a UV continuum slope of $\alpha = -2$ by fitting a power-law model of the form $f_{\nu} \propto \nu^{\alpha}$ to a \hst/Faint Object Spectrograph (FOS) spectrum of the AGN over the range 2200 to 3300 \AA. We fit a single power-law model to the SED of the AGN from the \texttt{GALFIT} decomposition over all seven photometric bands using the IDL-based least-squares fitting code \texttt{MPFIT} \citep{mpfit}, and measured a continuum slope $\alpha = -0.8 \pm 0.2$. We also independently fit power-law models to the SED in the UV and from the optical to the IR, and measured a slope of $-1.5 \pm 0.6$ between F275W and F336W, which is consistent with \citet{f&h&s1993}, and a measured a slope of $-0.3 \pm 0.4$ between F438W and F153M. However, our observations only represent a single temporal snapshot of the state of the AGN, and past observations have shown that the optical continuum shape of the AGN sometimes undergoes rapid changes. \citet{lira1999} found that the optical continuum slope changed from $-2$ to $0$ over a period of just six months between July 1996 and January 1997. The SED of the AGN from the \texttt{GALFIT} decomposition has an irregular shape, and the slope between the \emph{B}, \emph{V}, and \emph{I} bands is not smooth, suggesting that the fits are strongly affected by degeneracy between the AGN and compact cluster component. The photometric uncertainties from the \texttt{GALFIT} decomposition are quite large ($\sim 0.4$ mag for point source component and $\sim 0.2$ mag for the \sersic\ component in F814W), and improving this decomposition will require imaging of the NC with higher spatial resolution than what is currently available. 

\subsection{NGC 7793}
\indent At a distance of 3.4 Mpc \citep{piet2010}, the nearly face-on galaxy NGC 7793 is one of the brightest galaxies in the Sculptor Group. \citet{walcher2005} measured a stellar velocity dispersion of 24.6 \kms\ and a mass of $10^{6.89}$ \msun\ for the NC. NGC 7793 was included in the \citet{boker2002} \hst/WFPC2 NC imaging survey, where an \emph{I}-band absolute magnitude of $-13.64 \pm 0.03$ mag was measured. Correcting for the slightly different distances used, this corresponds to a WFC3 F814W absolute magnitude of $-13.61 \pm 0.03$ mag, which is in very good agreement with our measurement of $-13.59^{+0.06}_{-0.05}$ mag. We fit a flat sky to the host galaxy and a single \sersic\ component to the cluster in each band. The cluster has a larger extent in the UV than the IR, with an effective radius that monotonically decreases from $12.45^{+1.83}_{-1.83}$ pc in F275W to $7.94^{+2.73}_{-1.02}$ pc in F153M. The color gradient is obvious from inspection of the color composite image of the cluster shown in Figure \ref{master_color_images}. The \sersic\ index monotonically increases from $1.39^{+0.24}_{-0.27}$ in F275W to $3.29^{+1.18}_{-0.56}$ in F153M. Although a single \sersic\ component gives a reasonable fit in all bands, a closer inspection of the images reveals a ring-like gap in the stellar density, which is particularly prominent in the UV bands. This structure can also be seen in the residuals of the F814W fit (Figure \ref{img_fit_res_1}). 

\section{Conclusions}
\indent We have analyzed the structure of a sample of ten nuclear star clusters in bulgeless, late-type spiral galaxies by fitting PSF convolved 2D surface brightness models to \hst/WFC3 images of the clusters in seven wavebands ranging from the near-UV to the near-IR. We find that most of the NCs are described remarkably well by a single \sersic\ component, although some show clear evidence for multi-component substructure. NGC 2976 has a NC which consists of a compact, blue structure, which may be a young star cluster and an extended, red population. The NC in NGC 4244 contains a red, compact, spheroidal component and a blue, extended, flattened component. NGC 4395 hosts an unobscured AGN in its NC, which was modeled using a point source component. The main results of the surface brightness profile fits are as follows:  

\indent 1. The clusters exhibit a wide range of structural properties. In the F814W filter, absolute magnitudes range from $-11.20$ mag to $-15.02$ mag, effective radii range from $1.38$ to $8.28$ pc, \sersic\ indices range from $1.63$ to $9.73$, and axis ratios range from $0.57$ to $0.94$. We find a much wider range of \sersic\ indices than any previous study, which have generally measured $n\sim 1$ -- $3$. The highest \sersic\ indices we measure (above $~6$) serve as useful indicators of the shape of the wings of the NC profile, but do not provide information about the core concentration due to the effects of PSF blurring. 

\indent 2. We find evidence for spatially segregated stellar populations within some clusters. There are color gradients in six of the ten NCs in our sample, indicated by a dependence of the measured effective radius on wavelength. In M33, NGC 247, NGC 300 and NGC 4395, we find an increase in the effective radius with wavelength. This is consistent with general results found by \citet{g&b2014} for a large \hst/WFPC2 archival sample, and indicates the presence of a younger population which is more concentrated than the bulk of the NC stars. However, we find a general decrease in effective radius with wavelength in NGC 3621 and NGC 7793, which may indicate extended, circumnuclear star formation. For NGC 3621, the larger extent of the NC in UV bands may be due to scattered UV photons originating from the obscured Type 2 AGN. 

\indent 3. We also find a general trend of increasing roundness of the NCs at longer wavelengths, indicated by larger best-fit axis ratios in redder bands. There is also a correlation between the axis ratios of the NCs and their host galaxies. We conclude that blue disk structures aligned with the host galaxy plane are a common feature of NCs in late-type galaxies, but are more easily detected in galaxies that are close to edge-on.

\indent 4. A comparison of the measured colors of the NCs with SSP evolutionary tracks clearly shows that in general, a mix of young and old SSPs is needed to describe their stellar populations, consistent with previous spectroscopic studies of NCs \citep{rossa2006,walcher2006}. The mix of ages is only seen when considering colors which span the UV to the IR, and is not apparent from optical colors alone. The wide wavelength coverage of our data provides sensitivity to stellar populations with a mix of ages. 

\indent The surface brightness profiles presented in this paper will be used in future stellar population modeling and dynamical studies of the clusters. Recent studies have shown that degeneracies between age, metallicity and extinction can be broken by fitting SSP models to photometric SEDs of star clusters with UV to IR broadband coverage \citep{demeulenaer2014}. Our broad wavelength coverage will allow us to place tighter constraints on the star formation histories of the NCs than previous studies. By performing pixel by pixel SED fits to the surface brightness models presented in this paper, we can further search for population gradients within the NCs and constrain their stellar mass profiles. Stellar mass profiles, along with IFU stellar kinematic data of the clusters will be key ingredients in future dynamical studies of NCs, which may uncover a previously unseen population of intermediate-mass black holes within the clusters.  

\acknowledgments
Support for program GO-12163 was provided by NASA through a grant from the Space Telescope Science Institute, which is operated by the Association of Universities for Research in Astronomy, Inc., under NASA contract NAS 5-26555. LCH acknowledges support from the Kavli Foundation, Peking University, and the Chinese Academy of Science through grant No.\ XDB09030102 (Emergence of Cosmological Structures) from the Strategic Priority Research Program. MC acknowledges support from a Royal Society University Research Fellowship. This research has made use of the NASA/IPAC Extragalactic Database (NED) which is operated by the Jet Propulsion Laboratory, California Institute of Technology, under contract with the National Aeronautics and Space Administration. This publication makes use of data products from the Two Micron All Sky Survey, which is a joint project of the University of Massachusetts and the Infrared Processing and Analysis Center/California Institute of Technology, funded by the National Aeronautics and Space Administration and the National Science Foundation. Funding for the SDSS and SDSS-II has been provided by the Alfred P.\ Sloan Foundation, the Participating Institutions, the National Science Foundation, the U.S. Department of Energy, the National Aeronautics and Space Administration, the Japanese Monbukagakusho, the Max Planck Society, and the Higher Education Funding Council for England. The SDSS Web Site is http://www.sdss.org/. The SDSS is managed by the Astrophysical Research Consortium for the Participating Institutions. The Participating Institutions are the American Museum of Natural History, Astrophysical Institute Potsdam, University of Basel, University of Cambridge, Case Western Reserve University, University of Chicago, Drexel University, Fermilab, the Institute for Advanced Study, the Japan Participation Group, Johns Hopkins University, the Joint Institute for Nuclear Astrophysics, the Kavli Institute for Particle Astrophysics and Cosmology, the Korean Scientist Group, the Chinese Academy of Sciences (LAMOST), Los Alamos National Laboratory, the Max-Planck-Institute for Astronomy (MPIA), the Max-Planck-Institute for Astrophysics (MPA), New Mexico State University, Ohio State University, University of Pittsburgh, University of Portsmouth, Princeton University, the United States Naval Observatory, and the University of Washington. We would like to thank Chien Peng for helpful advice on using \texttt{GALFIT}. Finally, we would like the referee for carefully reading our manuscript and providing constructive comments which helped substantially improve the quality of the paper. 

\bibliography{ms}

\end{document}